\documentclass[12pt]{article}
\newcommand{\ft}[2]{{\textstyle\frac{#1}{#2}}}
\newsavebox{\uuunit}
\sbox{\uuunit}
    {\setlength{\unitlength}{0.825em}
     \begin{picture}(0.6,0.7)
        \thinlines
        \put(0,0){\line(1,0){0.5}}
        \put(0.15,0){\line(0,1){0.7}}
        \put(0.35,0){\line(0,1){0.8}}
       \multiput(0.3,0.8)(-0.04,-0.02){12}{\rule{0.5pt}{0.5pt}}
     \end {picture}}

\def\inbar{\vrule height1.5ex width.4pt depth0pt}
\def\IC{\relax\,\hbox{$\inbar\kern-.3em{\rm C}$}}
\newcommand{\Ka}{K\"ahler}
\newcommand{\three }{3-}
\newcommand{\kk}{k}
\newcommand{\ILambda}{I}
\newcommand{\JSigma}{J}
\newcommand{\XL}{X}
\newcommand{\FM}{F}
\def\ii{{\rm i}}
\def\ib{{\bar \imath}}
\def\jb{{\bar \jmath}}
\def\Im{{\rm Im ~}}
\def\Re{{\rm Re ~}}
\def\IR{\relax{\rm I\kern-.18em R}}
\def\IL{\relax{\rm I\kern-.18em L}}
\def\bfone{\relax{\rm 1\kern-.35em 1}}
\newsavebox{\zzzbar}
\sbox{\zzzbar}
  {\setlength{\unitlength}{0.9em}
  \begin{picture}(0.6,0.7)
  \thinlines
  \put(0,0){\line(1,0){0.6}}
  \put(0,0.75){\line(1,0){0.575}}
  \multiput(0,0)(0.0125,0.025){30}{\rule{0.3pt}{0.3pt}}
  \multiput(0.2,0)(0.0125,0.025){30}{\rule{0.3pt}{0.3pt}}
  \put(0,0.75){\line(0,-1){0.15}}
  \put(0.015,0.75){\line(0,-1){0.1}}
  \put(0.03,0.75){\line(0,-1){0.075}}
  \put(0.045,0.75){\line(0,-1){0.05}}
  \put(0.05,0.75){\line(0,-1){0.025}}
  \put(0.6,0){\line(0,1){0.15}}
  \put(0.585,0){\line(0,1){0.1}}
  \put(0.57,0){\line(0,1){0.075}}
  \put(0.555,0){\line(0,1){0.05}}
  \put(0.55,0){\line(0,1){0.025}}
  \end{picture}}
\newcommand{\Zbar}{\mathord{\!{\usebox{\zzzbar}}}}
\begin{document}
\begin{titlepage}
\begin{flushright}
KUL-TF-99/07\\
DFTT 6/99\\
IFUM-639-FT \\
hep-th/9902100
\end{flushright}
\vspace{.5cm}
\begin{center}
{\LARGE \bf  The 0-brane action in a general
\\ \vskip 0.2cm  $\mathbf{D=4}$ supergravity background}\\
\vfill
{\large Marco Bill\'o$^{1,2}$, Sergio Cacciatori$^{3}$, Frederik
Denef$^{2, +}$, Pietro Fr\'e$^1$,
\\
\vskip 0.2cm
Antoine Van Proeyen$^{2,\dagger}$
and  Daniela Zanon$^3$   } \\
\vfill
{\small
$^1$ Dipartimento di Fisica Teorica, Universit\'a di Torino, via P. Giuria 1,
I-10125 Torino, \\
 Istituto Nazionale di Fisica Nucleare (INFN) - Sezione di Torino, Italy \\
\vspace{6pt}
$^2$ Instituut voor Theoretische Fysica - Katholieke Universiteit Leuven
\\Celestijnenlaan 200D B--3001 Leuven, Belgium\\
\vspace{6pt}
$^3$ Dipartimento di Fisica, Universit\`a di Milano, via Celoria 16,
I-20133 Milano\\
and Istituto Nazionale di Fisica Nucleare (INFN) - Sezione di Milano, Italy\\
}
\end{center}
\vfill
\begin{center}
{\bf Abstract}
\end{center}
{\small We begin by presenting  the superparticle action in the
background of ${\cal N}=2$, $D=4$ supergravity coupled to $n$
vector multiplets interacting via an arbitrary special \Ka\
geometry. Our construction is based on implementing
$\kappa$-supersymmetry. In particular, our result can be
interpreted as the source term for ${\cal N}=2$ BPS black holes
with a finite horizon area. When the vector multiplets can be
associated with the complex structure moduli of a Calabi--Yau
manifold, our 0-brane action can then be derived by wrapping
3-branes around 3-cycles of the 3-fold. Our result can be
extended to the case of higher  supersymmetry; we explicitly
construct the $\kappa$ supersymmetric action for a superparticle
moving in an arbitrary ${\cal N}=8$ supergravity background with
$1/2$, $1/4$ or $1/8$ residual supersymmetry.}
 \vspace{2mm} \vfill \hrule width 3.cm
 {\footnotesize \noindent
 $^+$ Aspirant FWO, Belgium \\
 $^\dagger$ Onderzoeksdirecteur FWO, Belgium }
\end{titlepage}
\section{Introduction}
Recently, a lot of attention has been devoted \cite{Maldacena}
to the $AdS/CFT$ correspondence between
$D$-dimensional supergravity compactified on manifolds of the form
\begin{equation} \label{adscftp}
adS_{p+2} \, \times \, {\cal M}_{D-p-2}
\end{equation}
and conformal field theories on the  $(p+1)$-dimensional boundary
${\cal B}_{p+1}$ of anti~de~Sitter space.
In (\ref{adscftp}), ${\cal M}_{D-p-2}$ denotes a compact
$(D-p-2)$-dimensional
manifold whose geometrical structure largely
determines the specific properties of the boundary conformal field
theory.
\par
The situation envisaged by this correspondence arises in the context
of  BPS $p$-brane backgrounds with a suitable horizon behaviour.
These are classical solutions of $D$-dimensional supergravity that
interpolate  between two different vacua,
namely flat $D$-dimensional Minkowski space at
infinity and the product manifold (\ref{adscftp}) near the horizon \cite{GT}.
In addition, they are BPS states in the sense that they
preserve some amount of supersymmetry. As a consequence of this
there is a saturation of a bound relating the mass density to
the charge.
\par
Actually  $p$-branes are  solutions of {\em supergravity plus
sources} (see for instance \cite{stellelectur}): indeed, one  has to
supplement the bulk supergravity action with
the world-volume action of $p$-extended objects  carrying the
charges measured at spatial infinity.
 \par
The construction of these world-volume actions for $p$-branes is
mainly based on the principle that they should be $\kappa$-supersymmetric, namely that
there should be a suitable projection of the target space
supersymmetry that is promoted to a local fermionic symmetry on the
world-volume. Implementing $\kappa$-supersymmetry on the world-volume
puts the background supergravity fields on shell and requires that the superspace
Bianchi identities be satisfied \cite{kappa,castdauriafre,termoniaetal}.
Furthermore, by gauge-fixing the $\kappa$-symmetry, one halves the target
space fermionic coordinates providing the correct number of bosons
and fermions for a supersymmetric world-volume theory.
\par
When the $\kappa$-supersymmetric  action that describes the coupling
of the $p$-brane to generic supergravity backgrounds is known,
one can study its properties in  any given background solution.
Choosing an appropriate $\kappa$-gauge one derives
a consistent world-volume   field
theory   that inherits as global (super)symmetries the
(super)isometry group of the chosen background.
For instance, in the background (\ref{adscftp})
one can derive a {\em superconformal field theory} on the anti~de~Sitter
boundary ${\cal B}_{p+1}$ starting from the $p$-brane world-volume action \cite{conffadS}.
Indeed, in  this case the supersymmetric extension of the anti~de~Sitter
isometry group $SO(2,p+1)$ acts as the superconformal group on ${\cal B}_{p+1}$.
Such a construction has been used recently  to investigate the
properties of the {\em singleton conformal field theory} living on the
anti~de~Sitter boundary of M2 \cite{termoniaetal}
and \three  branes \cite{adS5S5}.
\par
 The BPS black-hole solutions of D=4 supergravities
 \cite{GH,FKS,RiccardoPietroReview} fit into
the above scheme as instances of $0$-branes.  However, the
$\kappa$-supersymmetric action for superparticles has been derived
so far only in the case of a pure $\mathcal{N}=2$ supergravity
background \cite{bhscmech}. The purpose of this paper is to extend
such a construction to more general supergravity backgrounds.
Firstly, we present the case of an arbitrary $\mathcal{N}=2$, $D=4$
background provided by supergravity coupled to $n$ vector
multiplets interacting via a generic  {\em special \Ka\ geometry}.
Secondly, we extend our result to ${\cal N}=8$ supergravity.
\par
The main new issue involved in our programme is the coupling of the
$0$-brane to the scalar fields. As we are going to see this coupling
occurs in a very simple and elegant way through a real function
\begin{equation}
\vert Z(\phi, p^\ILambda,q_\JSigma) \vert  = \frac{1}{\sqrt{2\nu }}
\, \left ( \widehat{Z}_{AB} \, \widehat{\overline{Z}}^{AB} \right
)^{1/2}\ , \label{funzia}
\end{equation}
where $\nu $ is the number of preserved supercharges,
sitting in front of the kinetic term. As will become apparent
in the following,  the real function (\ref{funzia}) is the modulus of
the largest skew eigenvalue of the field-dependent central charge
tensor $Z_{AB}$. The hat appearing in   (\ref{funzia}) denotes a
suitable projection operation that extracts the contribution from
the largest eigenvalue. The numbers $p^\ILambda,\,q_\JSigma$ are
the magnetic and electric charges of the black hole and $\phi$ are
the scalar fields. In the ${\cal N}=2$ case the scalars belong to
the vector multiplets, in the ${\cal N}=8$ case they belong to the
graviton multiplet, but the form (\ref{funzia}) of their coupling
to the superparticle action holds in both cases alike. This result
could be heuristically justified recalling that for a usual
particle the worldline action is multiplied by the particle mass
and that in the case of a BPS state the mass should be equal to
the central charge. The main difference is that in supergravity
the central charge is field dependent.
\par
Our result will be derived requiring $\kappa$-supersymmetry: it has a
general validity within ${\cal N}=2$ or ${\cal N}=8$ supergravity.
If the ${\cal N}=2$ supergravity model can be obtained from
the type~IIB superstring compactified on a Calabi--Yau (CY)
3-fold a geometric interpretation of our
$0$-brane action naturally arises: it corresponds to the
wrapping of the \three brane
world-volume action on a {\em supersymmetric cycle} of the
Calabi--Yau manifold.
\section{${\cal N}=2$ black holes solution}
The BPS black hole solutions of ${\cal N}=2$ pure supergravity have
been introduced in \cite{GH}. They have
been extended to the coupling with $n$ vector multiplets in
\cite{FKS}. The interplay between the BPS conditions and special \Ka\
geometry was found by \cite{SergioRenataGary}. In particular, in that
reference the relation of the black hole entropy with the
central charge and the so-called geodesic potential was clarified. These concepts
were further discussed in a vast literature; see, for example,
\cite{RiccardoPietroReview} for a review. The relevant features of these field
configurations are the following ones.
\begin{enumerate}
  \item The solutions are characterized by the vector of electric ($q_\JSigma$) and
  magnetic ($p^\ILambda$) charges carried by the black hole and by the values
  $z^i_\infty$
  of the moduli (scalar fields of the vector multiplet) at spatial infinity.
  \item The metric has a universal behaviour near the horizon $r=0$,
  where it approaches the Bertotti--Robinson metric \cite{berrobin}
\begin{equation}
ds^2_{\rm BR}=-\frac{1}{(G_N m_{\rm BR})^2} r^2 dt^2 +(G_N m_{\rm BR})^2 \, \frac{dr^2}{r^2}
+ (G_N m_{\rm BR})^2 \, \left( \sin^2(\theta) \, d\phi^2 + d\theta^2
\right)\ ,
 \label{bertrob}
\end{equation}
describing the geometry of $adS_2\times S^2$.
The parameter $m_{\rm BR}$, named the Bertotti--Robinson mass,
depends only on $p$ and $q$.
  \item The complex moduli fields $z^i$, starting at infinity from arbitrary
  values $z^i_\infty$, flow at the horizon to fixed values
$z^i_{\rm fix}$ that depend only on the quantized electric and magnetic
  charges $q$ and $p$.
\end{enumerate}
Let us briefly recall the basic ingredients of ${\cal N}=2$, $D=4$
supergravity and the way in which its BPS black hole solutions arise.
The bosonic part of the supergravity action is
(with $\kappa_4^2=8\pi G_N=1$)
\begin{eqnarray}
{\cal L}_{\rm B} & =&
\sqrt{-g}\Bigl  [- \ft12\,  R[g]
\, - \, g_{i\jb}(z,\bar z )\, \partial^{\mu} z^i \,
\partial_{\mu} \bar z^\jb\,
\nonumber\\
& & +\ft14 \,{\rm i} \,\left(
\bar {\cal N}_{\ILambda \JSigma} {\cal F}^{- \ILambda}_{\mu \nu}
{\cal F}^{- \JSigma \vert {\mu \nu}}
\, - \,
{\cal N}_{\ILambda \JSigma}
{\cal F}^{+ \ILambda}_{\mu \nu} {\cal F}^{+ \JSigma \vert {\mu \nu}} \right )
\, \Bigr ]\ .
\label{ungausugra}
\end{eqnarray}
The action is defined in terms of scalars $z^i$ and vectors
 $A^\ILambda_\mu $ whose field strength is the electric $2$-form
${\cal F}^\ILambda$:
\begin{equation}
{\cal F}^\ILambda= dA^\ILambda =\ft12
{\cal F}^\ILambda_{\mu\nu}dx^\mu \wedge dx^\nu =\ft12
(\partial_\mu A^\ILambda_\nu- \partial_\nu
A^\ILambda_\mu)dx^\mu \wedge dx^\nu\ .
 \label{FdA}
\end{equation}
Our convention for (anti)selfdual tensors is as follows:
\begin{equation}
{\cal F}^{\pm\,\ILambda}_{\mu \nu} = \ft{1}{2} \, \left({\cal F}^\ILambda_{\mu \nu} \pm {\rm i}
\ft{1}{2} \epsilon_{\mu \nu\rho\sigma} \, {\cal F}^{\ILambda\vert\rho\sigma}
\right)\ .
\label{fpm}
\end{equation}
The scalars are in a symplectic section of special geometry (see
\cite{DWVP,n2stand}):
\begin{equation}
V= \left( \matrix{ \XL^\ILambda \cr \FM_ \JSigma\cr} \right) \ ,
\end{equation}
where $e^{-{\cal K}/2}\XL^\ILambda(z^i)$, $e^{-{\cal K}/2}\FM_\JSigma(z^i)$
are  holomorphic, and ${\cal
K}$ is the \Ka\ potential. The covariant derivative of this section,
\begin{equation}
U_i = \nabla_i V = \left(\partial_i + \ft{1}{2} \,\partial_i{\cal K}
\right) V \equiv \left( \matrix { f^\ILambda_i \cr h_{\JSigma \vert i}
\cr } \right)\ ,
\label{fihi}
\end{equation}
defines $f^\ILambda_i$ and $h_{\JSigma \vert i}$, which in turn are
sufficient to determine the metric on the scalar manifold $g_{i\jb}$
and the kinetic matrix ${\cal N}$ in (\ref{ungausugra}):
\begin{eqnarray}
&& g_{i\jb}\,  =\ii \left(f_i^\ILambda \bar h_{\ILambda|\jb}-
h_{\ILambda|i} \bar f_\jb^\ILambda \right)~,
\nonumber\\
&&
{\bar \FM}_\JSigma = {\bar {\cal N}}_{\JSigma\ILambda} \, {\bar \XL}^\ILambda
\quad , \quad   h_{\JSigma \vert i} =  {\bar {\cal N}}_{\JSigma\ILambda}
f^{\ILambda}_{i}\ .
\label{scriptN}
\end{eqnarray}
The field equations of the vector fields involve the magnetic field
strengths
\begin{equation}
{\cal G}^{-}_{\JSigma \vert \mu \nu} = {\bar {\cal N}}_{\JSigma\ILambda}
{\cal F}^{-\ILambda}_{\mu \nu}\ ;\qquad
 {\cal G}^{+}_{\JSigma \vert \mu \nu} = {\cal N}_{\JSigma\ILambda}
{\cal F}^{+\ILambda}_{\mu \nu}\ ;\qquad
{\cal G}_{\JSigma \vert \mu \nu}={\cal G}^{-}_{\JSigma \vert \mu
\nu}+{\cal G}^{+}_{\JSigma \vert \mu \nu}\ ,
\label{Gpm}
\end{equation}
defined by the variation of the Lagrangian with respect to the
electric field strengths.
The electric and magnetic charges of the black hole are
defined by the formulae:
\begin{equation}
q_\ILambda  \equiv 
\frac{1}{4 \pi\kk} \,
\int_{S^2_\infty} \,{\cal G}_\ILambda
\ ;\qquad
p^\JSigma  \equiv
\frac{1}{4 \pi\kk} \,
\int_{S^2_\infty} \,{\cal F}^{\JSigma}\ .
\label{holmag}
\end{equation}
Note that $p$ and $q$ are not necessarily integer. Indeed, the Dirac
quantization condition gives that for two objects with charges $(p,q)$
and $(\tilde p,\tilde q)$
\begin{equation}
  8\pi \kk^2 \left(\tilde q_\ILambda\,p^\ILambda-\tilde p^\ILambda\, q_\ILambda
  \right)   \in \Zbar
\label{Diracqu}
\end{equation}
To have integer $p$ and $q$, allowing 1 as the lowest value,
we thus have to take $\kk=1/\sqrt{8\pi }$. We
will, however, leave $\kk$ arbitrary to facilitate comparison with
other papers.
\par
The black hole configuration solves the field equations from
the action (\ref{ungausugra}) and also satisfies the BPS
conditions, which are just the
statement that the supersymmetry transformations of the fermionic
fields, the gravitino $\psi_\mu^A$ and the gaugino $\lambda^{iA}$,
are zero:
\begin{eqnarray}
0 &=& {\cal D}_{\mu}\,\epsilon _A(x)\,
-\ft{\ii}{2} \epsilon_{AB} \, T^-_{\mu \nu}\, \gamma^{\nu}\epsilon^B(x)~,
 \nonumber \\
0 &=&
   \nabla_ {\mu}\, z^i  \, \gamma^{\mu} \epsilon^A(x)
-\ft12G^{-i}_{\mu \nu} \gamma^{\mu \nu} \epsilon_B(x) \epsilon^{AB}~,
\label{kigaug}
\end{eqnarray}
where ${\cal D}_{\mu}$ contains spin connection and \Ka\ connection.
The graviphoton  and matter field strengths, appearing in the above
transformations, are
\begin{eqnarray}
T^{-}_{\mu \nu } &=& \FM_\JSigma \,
{\cal F}^{-\JSigma}_{\mu \nu} - \XL^\ILambda \, {\cal G}^{-}_{\mu
\nu\vert \ILambda}~,\nonumber\\
G^{i-}_{\mu\nu} &=& - g^{i\jb}
 \bar f^\JSigma_\jb
\left (\Im  {\cal N}\right)_{\JSigma\ILambda}
\Bigl ( {\cal F}^{\ILambda -}_{\mu\nu}
\Bigr )~.\label{G-def}
\end{eqnarray}
BPS black holes are solutions of (\ref{kigaug}) with
  a supersymmetry parameter $\epsilon_A(x)$ of the form \cite{Moore}:
\begin{equation}
\gamma_0\epsilon_A (x)=\pm \frac{Z}{|Z|}\,\epsilon_{AB} \epsilon^B(x)\ ,
\label{quispi}
\end{equation}
where
\begin{equation}
Z(z,\bar{z},{p},{q}) =
\frac{1}{4\pi\kk} \,  \int_{S^2} \, T^{-} =
 \FM_\JSigma \, p^\JSigma - \XL^\ILambda \, q_\ILambda
\label{zentrum}
\end{equation}
is the central charge.
The projection (\ref{quispi}) halves the number
of components of the spinors, and it is the same projection, as we will
see, satisfied by the $\kappa$ symmetry parameter.
\par
In general the BPS black hole solution depends on radial functions
$U(r)$ and $z^i(r)$, to be determined below.  The 
metric is
\begin{equation}
ds^2 =- e^{2U(r)} \, dt^2 + e^{-2U(r)} \, d{\vec x}^2
\ ,\hskip 0.3cm (r^2 = {\vec x}^2)
\label{ds2U}
\end{equation}   admitting
$\IR \, \times \, SO(3)$  as the isometry group.
The spatial parts of the field strengths are
\begin{equation}
\label{flambda}
{\cal F}_s^{\ILambda}\,=\,\frac{\kk p^{\ILambda}}{2r^3}\epsilon_{ijk}
x^i dx^j\wedge dx^k\ ;\qquad
{\cal G}_{s,\ILambda}\,=\,\frac{\kk q_{\ILambda}}{2r^3}\epsilon_{ijk}
x^i dx^j\wedge dx^k
\end{equation}
\par
With this parametrization, the explicit form of the first-order BPS
conditions (\ref{kigaug}) is
\begin{eqnarray}
\frac{dz^i}{dr}\, &=&\pm \kk\left(\frac{e^{U(r)}}{r^2}\right) g^{i\jb}
\partial_\jb |Z(z,\bar{z},{p},{q})| \label{zequa}~,\\
\frac{dU}{dr}\, &=&\,\pm \kk\left(\frac{e^{U(r)}}{r^2}\right)
|\FM_\JSigma {p}^\JSigma-
\XL^\ILambda {q}_\ILambda|\,=\,
\pm \kk\left(\frac{e^{U(r)}}{r^2}\right)|Z(z,\bar{z},{p},{q})|\ .
\label{Uequa}
\end{eqnarray}
Only the upper sign is physical as argued in \cite{FredAttractor}.
\par
The first-order equations~(\ref{zequa}) have a fixed point, which we
put at $r=0$, by the condition $\nabla_\jb\bar{Z}(z,\bar{z},{p},{q})=0$.
This defines the {\em fixed} values $z^i_{\rm fix}$, and
correspondingly the fixed value of the central charge
\begin{equation}\label{Zfix}
  Z(p,q)=Z(z_{\rm fix},\bar{z}_{\rm fix},{p},{q})\ .
\end{equation}
The fixed point is the horizon. Indeed, in the vicinity of the fixed point
the differential equation for the metric becomes:
\begin{equation}
\frac{dU}{dr}=\kk\frac{|Z(p,q)|}{  \, r^2} \, e^{U(r)}\ ,
\end{equation}
which has the asymptotic solution:
\begin{equation}
\exp[-U(r)]\, {\stackrel{r \to 0}{\longrightarrow}}\,
\mbox{constant} + \kk\frac{|Z(p,q)|} { r}\ .
\label{approxima}
\end{equation}
Hence, near $r=0$   the metric (\ref{ds2U})
becomes of the Bertotti--Robinson type (\ref{bertrob}),
with Bertotti--Robinson  mass:
\begin{equation}
 m_{\rm BR} = 8\pi \kk \vert  {Z(p,q)}  \vert\ .
\label{brmass}
\end{equation}

\section{The $\kappa$ supersymmetric action of a $0$-brane }
In this section we derive a simple formula for the worldline action
of a $0$-brane moving in the background geometry
provided by a generic {\em on-shell} field configuration of ${\cal N}=2$
supergravity. In particular, our action describes
the 0-branes corresponding to the black hole solutions
discussed above.
\par
To write the worldline action we use Polyakov first-order
formalism. Hence we introduce the worldline  {\it einbein} 1-form
\begin{equation}
   e= e_ \tau  \, d\tau
   \label{einbein}
\end{equation}
and the auxiliary $0$-forms $ \Pi^a $ that, on shell, become the
worldline components of the spacetime supervielbein, satisfying
the supergravity Bianchi identities \cite{kappa,castdauriafre,termoniaetal}:
\begin{equation}
V^a = \Pi^a \, e\ .
\label{piae}
\end{equation}
We write the following ansatz for the worldline action:
\begin{equation}
S_{\rm wl} =4\pi\kk \left\{ \int_{M_1} \, R(z,\bar z) \, \left [ \left (-\Pi^a \, V^b
+ \ft{1}{2} \, \Pi^a \, \Pi^b  \, e \right) \, \eta_{ab} +\ft{1}{2}
\, e \right]+ \int _{M_2} \, \left( p^\JSigma \, {\cal G}_\JSigma - q_\ILambda \,
{\cal F}^\ILambda \right)\right\} \ ,
\label{0act}
\end{equation}
where $(p^\JSigma, q_\ILambda)$ is the vector of
integer magnetic and electric charges (\ref{holmag}), while
$R(z, \bar z)$ is a {\em real} function of the scalar fields that we
have to determine in such a way that the action is
$\kappa$-supersymmetric. Furthermore, in the Wess--Zumino (WZ) term we
introduced an integral over a two-dimensional manifold $M_2$ whose
boundary is the worldline $ M_1 = \partial \, M_2$.
The normalization of the WZ term is fixed by (\ref{holmag}) and the
vector equation of motion.
\par
The action (\ref{0act}) is to be varied independently  with respect to:
\begin{enumerate}
\item The auxiliary 0-forms $\Pi^a$. Such a variation yields the
identification (\ref{piae}) as a field equation.
\item The einbein $e$. Such an equation yields:
\begin{equation}
\eta_{ab} \, \Pi^a \, \Pi^b =-1\ ,
\label{idepi}
\end{equation}
which is the intrinsic way of stating that the worldline
metric is the induced one from the target spacetime metric. Indeed
equation~(\ref{idepi}) can be read in the following way:
\begin{equation}
\eta_{ab} \, V^a_\mu \, V^b_\nu \, \frac{d x^\mu }{d\tau}
\, \frac{d x^\nu }{d\tau} h^{\tau\tau} =1\ ,
\label{idever}
\end{equation}
where
\begin{equation}
h^{\tau\tau} = \frac{1}{h_{\tau\tau}} \quad ; \quad  h_{\tau\tau}=
-e_\tau e_\tau \label{defh}
\end{equation}
denotes the contravariant worldline metric.
\item The target superspace coordinates $x$ and $\theta$.
This yields the second-order field equations.
\end{enumerate}
To discuss $\kappa$-supersymmetry we need the ordinary ${\cal N}=2$
supersymmetry transformation rules of the bosonic fields
\cite{dWLVP,n2stand}:
\begin{eqnarray}
\delta\,V^a_{\mu}&=& \,\bar \epsilon^A\,\gamma^a\,{\psi}_{A
\mu} +\,\bar \epsilon_A\,\gamma^a\,{\psi}^A _\mu~,\nonumber\\
\delta\,A^\ILambda _{\mu}&=&
2 \bar \XL^\ILambda \bar \psi _{A\mu} \epsilon _B
\epsilon^{AB}\,+2\XL^\ILambda\bar\psi^A_{\mu}\epsilon^B \epsilon_{AB}
-\left(f^{\ILambda}_i \,\bar {\lambda}^{iA}
\gamma _{\mu} \epsilon^B \,\epsilon _{AB} +
{\bar f}^{\ILambda}_\ib \,\bar\lambda^\ib_A
\gamma _{\mu} \epsilon_B \,\epsilon^{AB}\right)~,\nonumber\\
\delta\,z^i &=& \bar{\lambda}^{iA}\epsilon _A~,\nonumber\\
\delta\,z^\ib&=& \bar{\lambda}^\ib_A \epsilon^A\ .
\label{veeltrasf}
\end{eqnarray}
In the above formulae $\bar {\lambda}^{iA}$ and
$\bar\lambda^\ib_A$ denote the two chiral projections of the
(conjugate) gaugino field.
In addition to $A^\ILambda_\mu$ we need to introduce a dual magnetic potential
$B_{\JSigma\vert \mu }$ whose field strength  is ${\cal G}_{\JSigma}$:
\begin{equation}
{\cal G}_{\JSigma}=d B_\JSigma=\ft12
{\cal G}_{\JSigma|\mu\nu}dx^\mu \wedge dx^\nu =\ft12( \partial_\mu B_{\JSigma|\nu}- \partial_\nu
B_{\JSigma|\mu})dx^\mu \wedge dx^\nu\ .
 \label{GdB}
\end{equation}
By consistency, the supersymmetry transformation rule of the dual
magnetic potential $B_\ILambda$ is defined as
\begin{eqnarray}
\delta\,B_{\ILambda \vert\mu}&=&
2 \bar \FM_\ILambda \bar \psi _{A\mu} \epsilon _B
\epsilon^{AB}\,+2 \FM_\ILambda\bar\psi^A_{\mu}\epsilon^B \epsilon_{AB}
-\left(h_{\ILambda\vert i} \,\bar {\lambda}^{iA}
\gamma _{\mu} \epsilon^B \,\epsilon _{AB} +
{\bar h}_{\ILambda \vert\ib} \,\bar\lambda^\ib_A
\gamma _{\mu} \epsilon_B \,\epsilon^{AB}\right)\ .
\label{Btrasf}
\end{eqnarray}
\par
The $\kappa$ transformation is simply a supersymmetry
transformation where the supersymmetry parameter $ \epsilon _A$ =
$\kappa _A$ is projected on the $0$-brane through the following
equations:
\begin{eqnarray}
\kappa _A \, + \, \epsilon _{AB} \Pi^a \, \gamma _a \,
\kappa^B  e^{i\varphi} &= & 0~,  \nonumber\\
\kappa^A \,+ \, \epsilon^{AB} \Pi^a \, \gamma _a \,
\kappa _B  e^{-i\varphi} &= & 0\ ,
\label{proje}
\end{eqnarray}
where $\varphi$ is an appropriate phase that we  identify below.
Note that the above 0-brane projection on the $\kappa$-supersymmetry
parameter is identical in form to the condition  (\ref{quispi}) imposed on the
parameter of the supersymmetry preserved by the BPS black holes. This can be
easily checked by going to a static gauge where the
worldline time $\tau$ is identified with the coordinate time $t$.
\par
Now we prove the following statement:
{\em  the action (\ref{0act}) is $\kappa$-supersymmetric if the real
function is chosen as follows:
\begin{equation}
R(z, \bar z) = -2| Z(z,\bar z)|\ .
\label{Rfun}
\end{equation}}
We obtain the proof by direct verification
using the `$1.5$-order formalism'. This
means that we vary the action (\ref{0act}) only in the superspace coordinates $x$
and $\theta$ and after variation we implement the field equations of the
auxiliary fields $\Pi^a$ and $e$. \par
With this proviso there are  only  three relevant variations, namely that
of the vielbein $V^a$, that of the gauge fields $A^\ILambda, B_\JSigma$
and that of the scalars $z^i$, ${\bar z}^\ib$.
In this way we produce two kinds of terms, those containing the gravitino
  $\psi_A, \psi^A$ and those containing the gaugino
${\bar \lambda}^{iA}, {\bar \lambda}^\ib_A$. Such terms have to
cancel separately. The variation of the gauge fields contains both
type of terms, while the variation of the vielbein contains only
the gravitino and the
variation of the scalars contains only the gauginos (see
(\ref{veeltrasf}) and (\ref{Btrasf})).
Let us concentrate first on the gravitino terms. We obtain:
\begin{eqnarray}
\frac{1}{4\pi \kk}\delta_\kappa \, S_{\rm wl} & = & \int_{M_1} R(z,\bar z)\, \left[
 {\bar \psi}_A \, \gamma_a \kappa^A +
 {\bar \psi}^A \, \gamma_a \kappa_A  \right] \, \Pi^a \nonumber\\
 && -2 q_\ILambda \, \int_{M_1} \, \left( {\bar \XL}^\ILambda \,
 {\bar \psi }_A \, \kappa_B \, \epsilon^{AB} +
 {  \XL}^\ILambda \,
 {\bar \psi }^A \, \kappa^B \, \epsilon_{AB}  \right)\nonumber\\
 && +2 p^\JSigma \, \int_{M_1} \,  \left( {\bar \FM}_\JSigma \,
 {\bar \psi }_A \, \kappa_B \, \epsilon^{AB} +
 {  \FM}_\JSigma \,
 {\bar \psi }^A \, \kappa^B \, \epsilon_{AB}  \right)\ .
 \label{psivar}
\end{eqnarray}
We simplify the terms in (\ref{psivar}) that contain  the $\gamma_a$ matrix
by use of the projection property (\ref{proje}):
\begin{eqnarray}
 {\bar \psi}_A \, \gamma_a \kappa^A \, \Pi^a  &=&-
 e^{- {\rm i}\, \varphi}\,
 \epsilon^{AB}  {\bar \psi}_A \, \gamma_a \, \gamma_b \kappa_B \,
 \Pi^a \, \Pi^b =-e^{- {\rm i}\, \varphi}\,
 \epsilon^{AB}  {\bar \psi}_A \,  \kappa_B \,
 \Pi^a \, \Pi^b \, \eta_{ab} \nonumber\\
 &=& e^{- {\rm i}\, \varphi}\,
 \epsilon^{AB}  {\bar \psi}_A \, \kappa_B \,  \nonumber\\
 {\bar \psi}^A \, \gamma_a \kappa_A \, \Pi^a & =&-
 e^{{\rm i}\, \varphi}\,
 \epsilon_{AB}  {\bar \psi}^A \, \gamma_a \, \gamma_b \kappa^B \,
 \Pi^a \, \Pi^b =-
 e^{{\rm i}\, \varphi}\,
 \epsilon_{AB}  {\bar \psi}^A \,  \kappa^B \,
 \Pi^a \, \Pi^b \, \eta_{ab}\nonumber \\
 &  = & e^{{\rm i}\, \varphi}\,
 \epsilon_{AB}  {\bar \psi}^A \, \kappa^B\ ,
 \label{reduz}
\end{eqnarray}
where we have also used  the first-order equation~(\ref{idepi}).
Inserting  (\ref{reduz}) into (\ref{psivar}) we see that all
the ${\bar \psi}_A, {\bar \psi^A}$ terms cancel if
\begin{eqnarray}
{\bar Z} \equiv  \left( p^\JSigma \,{\bar  \FM}_\JSigma - q_\ILambda \, {\bar \XL}^\ILambda \right) & = &
-\ft12 R (z, \bar z ) \, e^{-{\rm i}\, \varphi}~,\nonumber\\
Z \equiv \left( p^\JSigma \,{   \FM}_\JSigma - q_\ILambda \, {\XL}^\ILambda \right) & = &
-\ft12 R (z, \bar z ) \, e^{{\rm i}\, \varphi}~.
\end{eqnarray}
Hence we conclude that
\begin{equation}
R(z,\bar z) =   - 2\, \vert\,  Z  \, \vert \quad ; \quad \varphi(z,\bar z) =
\ft{1}{2}{\rm i}
\, \log\frac{\bar Z}{Z}~.
\label{pisel}
\end{equation}
\par
At this point the action is completely fixed and no further
parameters can be adjusted. The very non-trivial check of
$\kappa$-supersymmetry is that the same choices
(\ref{pisel}) needed to cancel the
gravitino  terms guarantee
the cancellation of gaugino terms as well. To verify this,
we need to consider the variation of the real
function $R(z, \bar z)$. We obtain:
\begin{eqnarray}
\delta_\kappa R(z,\bar z)& = & - \frac{1}{\vert Z \vert }\,
\left[ {\bar Z} \, \nabla_i Z \, \delta z^i + Z \, \nabla_\ib
{\bar Z} \, \delta {\bar z}^\ib \right] \nonumber\\
& = & - e^{-{\rm i}\varphi} \, \left(
p^\ILambda \, h_{\ILambda \vert i} - q_\JSigma \, f^\JSigma_i \right) \,
{\bar \lambda}^{iA} \, \kappa_A  - e^{{\rm i}\varphi}\left(
p^\ILambda \, {\bar h}_{\ILambda \vert \ib} -
q_\JSigma \, {\bar f}^\JSigma_\ib \right) \,
{\bar \lambda}^\ib_A \, \kappa^A \label{lavario}~.
\end{eqnarray}
Note that in deriving (\ref{lavario}) we have used the property that
the central charge is covariantly holomorphic $\nabla_\ib Z =
0$.
Using (\ref{lavario}) we conclude that at the level of the gaugino terms the
$\kappa$-supersymmetry variation of the worldline action is as
follows:
\begin{eqnarray}
\frac{1}{4\pi\kk}\delta_\kappa \, S_{\rm wl}  & =  & \int_{M_1}  \left[
\left(p^\ILambda \, h_{\ILambda \vert i} - q_\JSigma \, f^\JSigma_i
\right) \, {\bar \lambda}^{iA} \, \kappa_A \, e^{-{\rm i}\varphi}
+\left(
p^\ILambda \, {\bar h}_{\ILambda \vert \ib} -
q_\JSigma \, {\bar f}^\JSigma_\ib \right) \,
{\bar \lambda}^\ib_A \, \kappa^A \, e^{{\rm i}\varphi}\right]
\, \Pi^a \, V^b \, \eta_{ab}
\nonumber\\
&  -& \left[ \left(p^\ILambda \, h_{\ILambda \vert i} - q_\JSigma \, f^\JSigma_i
\right) \,
{\bar \lambda}^{iA} \,\gamma_a \kappa^B \, \epsilon_{AB} +
\,\left(
p^\ILambda \, {\bar h}_{\ILambda \vert \ib} -
q_\JSigma \, {\bar f}^\JSigma_\ib \right) \,
{\bar \lambda}^\ib_ A \,\gamma_a \kappa_B \, \epsilon^{AB} \,
\right] \, V^a~.
\label{lamka}
\end{eqnarray}
Using the projection property (\ref{proje}) of the $\kappa$-supersymmetry
parameter we can reduce the terms containing the $\gamma_a$ matrix
in complete analogy to (\ref{reduz}):
\begin{eqnarray}
 {\bar \lambda}^{iA} \, \gamma_a
\, \kappa^B \, \epsilon_{AB} \, V^a &=&
- e^{-{\rm i}\varphi} \,   {\bar \lambda}^{iA} \,\gamma_a
 \, \epsilon^{BC} \, \Pi^b \, \gamma_b \, \kappa_C \,
 \epsilon_{AB} \, V^a =e^{-{\rm i}\varphi} {\bar \lambda}^{iA}
 \kappa_A \, V^a   \Pi^b \, \eta_{ab}~, \nonumber\\
{\bar \lambda}^\ib_{A} \, \gamma_a
\, \kappa_B \, \epsilon^{AB} \, V^a &=&
- e^{{\rm i}\varphi} \,   {\bar \lambda}^\ib_{A} \,\gamma_a
 \, \epsilon_{BC} \, \Pi^b \, \gamma_b \, \kappa^C \,
 \epsilon_{AB} \, V^a =e^{{\rm i}\varphi} {\bar \lambda}^\ib_{A}
 \kappa^A \, V^a   \Pi^b \, \eta_{ab}~.
 \label{riduc}
\end{eqnarray}
Inserting the identity (\ref{riduc}) into (\ref{lamka}) we verify that
all the gaugino terms cancel identically. This concludes the proof of
our statement.
\par
Now that we have shown that the action (\ref{0act}) is supersymmetric
with the choice (\ref{pisel}), we can use the first-order field
equations~(\ref{idepi}) and (\ref{piae}) to recast the action in
second-order form. To do this explicitly at the level of both the
fermionic and bosonic coordinates we would need an explicit
parametrization of the spacetime supervierbein $V^a$ in terms of both $x$
and $\theta$. For the bosonic part we obtain
\begin{equation}\label{S2ndorder}
  S=4\pi\kk\left[ -2\int_{M_1}  |Z| \sqrt{-h_{\tau\tau}} d\tau +
   \int_{M_2} \, \left( p^\JSigma \, {\cal G}_\JSigma - q_\ILambda \,
{\cal F}^\ILambda \right)\right] \ ,
\end{equation}
where the induced metric $h_{\tau\tau}$, defined by (\ref{defh}), is
given in (\ref{idever}). Note that the mass we find here as a source
term for gravity agrees with (\ref{brmass}).

\section{Derivation from \three  brane wrapping}
The $0$-brane action constructed in the previous section has a fully
general validity. Indeed, as we emphasized above,
its $\kappa$ supersymmetry relies only on the general identities of
special \Ka\ geometry, irrespective of whether the vector
multiplet complex scalars  $z^i, \bar z^\ib$ can be
interpreted as {\em moduli} of a Calabi--Yau manifold or not.
\par
In the case where  vector multiplets are associated with
 {\em complex structure  moduli} of a Calabi--Yau
3-fold ${\cal M}_{\rm CY}$, giving the compactification of type~IIB string theory
from 10 to 4 dimensions, then our $0$-brane action admits a
geometrical interpretation as the result of wrapping \three branes
along suitable $3$-cycles. For simplicity, we restrict our attention to
the bosonic action (\ref{S2ndorder}). The full
identification is guaranteed by $\kappa$-supersymmetry that has
already been proven.
\par
We start from the bosonic world-volume action of a \three  brane in type~IIB string
theory; for our purpose we can limit ourselves to its `Nambu--Goto'
kinetic term and its coupling to the Ramond--Ramond 4-form, namely
\begin{equation}\label{D3braneAction}
  S_{3}=-\frac{T_3}{\kappa_{10}}\int_{M_4} d^4\xi \sqrt{-\det
  h_{mn}
  }+ \mu_3\int_{M_5} F_5^+\ .
\end{equation}
Here $h_{mn}(\xi)$ is the induced metric
\begin{equation}\label{inducedmetric}
h_{mn}=\frac{\partial X^M}{\partial\xi^m} G_{MN}
\frac{\partial X^N}{\partial\xi^n} \ ,
\end{equation}
where $G_{MN}$ is the 10-dimensional metric, and the fields $X^M(\xi)$
describe the embedding of the brane in the 10-dimensional space.
The tension of the \three brane is $T_3=\sqrt{\pi}$ and it is related to
its RR charge
by the BPS condition $\mu_3 = \sqrt{2} T_3$. Further, $\kappa_{10}$ is the
10-dimensional gravitational coupling constant which appears in
the  Einstein--Hilbert part of the `bulk' type~II
supergravity action as
\begin{equation}\label{IIBact}
  S_{\rm IIB}=-\frac{1}{2\kappa_{10}^2}\int d^{10}X \, \sqrt{G}\,R(X)+\ldots \ .
\end{equation}
We wrote the Wess--Zumino
term as an action on a five-dimensional manifold
$M_5$, having the \three  brane world-volume $M_4$ as its boundary.
\par
We compactify the theory to 4 dimensions on a Calabi--Yau space;
accordingly, we decompose the 10-dimensional coordinates as
$X^{M}=\left( x^\mu,y^\alpha,\bar y^{\bar\alpha}\right)$. The metric
has a direct product form: the only non-zero entries
are $G_{\mu\nu}$, the 4 dimensional metric, and $G_{\alpha \bar\beta}$,
the metric on the CY space. Notice that upon compactification,
the integral over the Calabi--Yau space contributes in (\ref{IIBact})
with a factor
\begin{equation}\label{VCY}
  V_{\rm CY}=\int_{{\cal M}_{\rm CY}}\ii \, d^3 y \, d^3 {\bar y} \, \left|\det
G_{\alpha \bar \beta}\right|\ ,
\end{equation}
the volume of the CY manifold. Therefore, the effective gravitational
constant in 4 dimensions, that, as usual,
we have conventionally taken to be 1, is
\begin{equation}\label{kappa410}
  \kappa_4=\frac{\kappa_{10}}{V_{\rm CY}^{1/2}}=1\ .
\end{equation}
This will be important below.
\par
We consider the case in which the \three brane is wrapped on a non-trivial
3-cycle of the internal CY manifold. Then
the embedding of the brane is as follows:
\begin{equation}\label{wrapping}
x^\mu (\xi^0)\ ;\qquad y^\alpha(\xi^i)\ ;\qquad \bar
y^{\bar\alpha}(\xi^i)\ ,
\end{equation}
having split the coordinates on the world-volume as $\xi^m=(\xi^0,\xi^i)$,
with $i=1,2,3$.
\par
We now consider separately the kinetic terms and WZ-terms.
\paragraph{Wrapping the kinetic term.}
 Although it
is very simple, we do not write the explicit form of the first-order \three brane
action since we can reach our conclusion by working directly in the second-order
formalism.
The second-order bosonic action of the \three brane (\ref{D3braneAction}) is given
in terms of the induced metric (\ref{inducedmetric}). Due to the direct product structure
of the metric $G$, the metric $h_{mn}$ is block diagonal, and $\det(-h_{mn})$
is a product of $-h_{00} = -\dot x^\mu G_{\mu\nu}\dot x^\nu$
times the determinant of the $3\times 3$ matrix
\begin{equation}
\label{comph}
h_{ij}= 2\frac{\partial y^\alpha}{\partial \xi^{(i}}
\frac{\partial y^{\bar \beta}}{\partial \xi^{i)}}\,
G_{\alpha\bar\beta}~,
\end{equation}
that is the pullback of the CY \Ka\ metric $G_{\alpha\bar\beta}$
on the 3-cycle.
\par
As a next step we show that the integral on a $3$-cycle ${\cal C}^3$
of the `spatial' part of the kinetic action (\ref{D3braneAction})
is proportional (through a constant)
to the modulus of the central charge
\begin{equation}\label{tobeproven}
  \int_{{\cal C}^3} d^3\xi \sqrt{\det h_{ij}} = 8\sqrt{\pi}\kk\,
  V_{\rm CY}^{1/2}\, |Z|\ ,
\end{equation}
if ${\cal C}^3$ is a {\em supersymmetric} cycle~\cite{BBS}.
A supersymmetric 3-cycle
of a CY is described by an embedding $f: \ {\cal C}^3 \longrightarrow {\cal M}_{\rm CY}$
satisfying two conditions.
\begin{enumerate}
\item The cycle is a Lagrangian
submanifold, namely the pull-back of the \Ka\ 2-form $J$ vanishes:
\begin{equation}
0=f^*J=\frac{\rm i}{2\pi} G_{\alpha\bar\beta}
\frac{\partial y^\alpha}{\partial \xi^i}
\frac{\partial y^{\bar\beta}}{\partial \xi^j} d\xi^i\wedge d\xi^j~.
\label{nr1}
\end{equation}
\item
Introduce the function $\phi$ as follows:
\begin{equation}
f^*\Omega= \Omega_{\alpha\beta\gamma}\frac{\partial y^\alpha}{\partial \xi^i}
\frac{\partial y^\beta}{\partial \xi^j}\frac{\partial y^\gamma}{\partial
\xi^k}\, d\xi^i\wedge d\xi^j \wedge d\xi^k\equiv 6\phi\, d^3\xi~,
\label{nr2}
\end{equation}
where $\Omega$ is the unique holomorphic $3$-form on the CY space,
and in our conventions
$d\xi^i\wedge d\xi^j \wedge d\xi^k=\epsilon^{ijk}d^3\xi$.
Then {\em the phase of $\phi$ has to be  constant}. As will become evident,
this is a minimal volume condition.
\end{enumerate}
It is not guaranteed {\it a~priori} that supersymmetric representatives
exist in any cohomology class; their existence
may thus restrict the allowed values of electric and magnetic
charges. The latter specify indeed, as we shall see below, the cohomology class
of the 3-cycle.
\par
In order to compute $\det h_{ij}$ we need to consider
 the induced metric on the world-volume
\begin{equation} \label{risulat}
  \det h_{ij} = 8\left|\det (\partial_i y^\gamma )
\right|^2 \det G_{\alpha\bar\beta}\ ,
\end{equation}
using  the definition (\ref{comph}),
combined with the first of the two conditions defining a
supersymmetric cycle, equation~(\ref{nr1}).
Since $\Omega$ is covariantly constant on the CY 3-fold, its norm
\begin{equation}\label{dettog}
  \|\Omega\|^2 \equiv \ft16 G^{\alpha \bar\alpha '}G^{\beta\bar\beta'}G^{\gamma\bar
  \gamma'}\Omega_{\alpha \beta\gamma}\bar\Omega_{\bar\alpha '\bar\beta'\bar \gamma'}
 = (\det G_{\alpha\bar\beta})^{-1}|\Omega_{123}|^2
\end{equation}
is a constant.
{}From the above relation we can obtain the expression of
the determinant of the Calabi--Yau \Ka\ metric and insert it
into (\ref{risulat}). We
find
\begin{equation}\label{urone}
  \det \, h_{ij}= \frac{8}{\|\Omega\|^2}|\det (\partial_i y^\alpha
  )\Omega_{123}|^2= 8 \frac{|\phi|^2}{\|\Omega\|^2}\ ,
\end{equation}
where in the last step we have used
$\phi=\Omega_{123}\det (\partial_i y^\alpha )$, as follows from
 (\ref{nr2}).
\par
Next we recall that for the special \Ka\ geometry the \Ka\
potential ${\cal K}$ is defined as
\begin{eqnarray}
\label{mohica}
e^{-{\cal K}}=\int_{{\cal M}_{\rm CY}}\ii\Omega\wedge \bar\Omega= 36
\|\Omega\|^2\,V_{\rm CY} \ ,
\end{eqnarray}
with $V_{\rm CY}$ defined in (\ref{VCY}).
Here and in the following it is useful to introduce the
rescaled $(3,0)$ form
\begin{equation}\label{scaledOmega}
  \widehat{\Omega}=e^{{\cal K}/2}\Omega\ ,
\end{equation}
which thus satisfies $\int_{{\cal M}_{\rm CY}}\widehat{\Omega}\wedge
\bar{\widehat{\Omega}}=-\ii$.
Inserting (\ref{mohica}) into (\ref{urone}) we obtain
\begin{equation}
{1\over 2\sqrt{2}}\int_{{\cal C}^3} d^3\xi \sqrt{\det h_{ij} }=
\int_{{\cal C}^3} d^3\xi\frac{|\phi|}{\|\Omega\|}\,\geq\,
\frac{1}{\|\Omega\|}\left|\int_{{\cal C}^3} d^3\xi \phi\right|=
\frac{1}{6\|\Omega\|}\left|\int_{{\cal C}^3} f^*\Omega\right|=
V_{\rm CY}^{1/2}\left|\int_{{\cal C}^3}f^*\widehat{\Omega}\right|\ .
\end{equation}
It is now clear that the bound is saturated whenever the
phase of $\phi $ is constant over the 3-brane.
In virtue of condition (\ref{nr2}) this happens for the
supersymmetric cycles. Therefore, for these latter cycles we obtain indeed
(\ref{tobeproven}), since the central charge can be expressed as
\begin{equation}\label{veravera}
 \sqrt{8\pi}\kk\,Z= \int_{{\cal C}^3}f^*\widehat{\Omega}~.
\end{equation}
Equation~(\ref{veravera}) is established through the
following steps. First one recalls that
 the holomorphic symplectic section of special geometry is
provided by the periods  of the holomorphic
3-form along a homology basis:
\begin{equation}\label{periodi}
\left( \matrix { \XL^\ILambda \cr \FM_\JSigma \cr }\right) = \left(\matrix{
\int_{A^\ILambda } \, \widehat{\Omega} \cr  \int_{B_\JSigma } \, \widehat{\Omega}
\cr}\right)\ .
\end{equation}
Then one interprets the electric and magnetic charges of the
black hole as the components of the cycle ${\cal C}^3$ in the same homology basis:
\begin{equation}
{\cal C}^3 = \sqrt{8\pi}\kk\left( -q_\ILambda \, A^\ILambda + p^\JSigma \, B_\JSigma
\right) \ .
\end{equation}
The overall factor is introduced because $p$ and $q$ have arbitrary
normalization. Considering (\ref{Diracqu}), we see that the minimal
value of $q$ and $p$ is $1/( \sqrt{8\pi}\kk)$, leading to the factor
above.
Comparing with (\ref{zentrum}), this leads immediately to
 (\ref{veravera}).
\par
In this way we have established that wrapping along a
supersymmetric 3-cycle the kinetic term of the \three brane we get
\begin{eqnarray}
  S_{\rm 3 ,Kin}&\equiv&-\frac{T_3}{\kappa_{10}}\int_{M_4} d^4\xi \sqrt{-\det
  h_{mn}}=-\frac{T_3}{\kappa_{10}}\int_{M_1} d\tau \sqrt{-h_{00}}\,
2\sqrt{2}\,
  V_{\rm CY}^{1/2}\, \sqrt{8\pi}\kk\, |Z| \nonumber\\
  &=&-8\pi \kk\int_{M_1}\, |Z|  \sqrt{-h_{00}}\,d\tau\ ,\label{resultkin}
\end{eqnarray}
which is the kinetic term of the $0$-brane in the form we have
discussed in the previous section.

\paragraph{Wrapping the Wess Zumino term.}
To obtain the WZ part of the $0$-brane action, we consider the
WZ action for the \three  brane action (\ref{D3braneAction}), written as an action on a
five-dimensional manifold $M_5$, having the \three brane world-volume as its
border. The Ramond--Ramond field strength $F_5^+$ is real and
self-dual. Using the results in (\ref{dualin10}), it contains upon
compactification  self-dual field strengths in 4 dimensions multiplied
by $(3,0)$ and $(1,2)$ forms and antiself-dual field strengths with
$(0,3)$ and $(2,1)$ forms. Considering \Ka\ weights and
symplectic invariance we fix the expression of the WZ action in terms
of the graviphoton and matter field strengths $T$ and $G^i$ defined in
(\ref{G-def}):
\newcommand{\prefactor}{\mu_3 b}
\begin{equation}\label{WZTG}
  S_{3 ,\rm WZ}=\prefactor\int_{M_5}\left[ \widehat{\Omega}^{(3,0)}(-{\rm i})T^+ +
  \widehat{\Omega}^{(1,2)}_{\bar\imath} 2 G^{+\,\bar\imath} + c.c.\right]\ .
\end{equation}
The factor $b$ should be chosen such that $b$ times the integral
gives $\sqrt{2\pi}\Zbar $, according to the quantization condition
(see e.g. (13.3.12) and (13.3.13) in \cite{Polchinski2}).
Here, $\widehat{\Omega}^{(3,0)}$ is the $(3,0)$ form which we
previously just denoted as $\widehat{\Omega}$, while
$\widehat{\Omega}^{(1,2)}_{\bar\imath}$ are a basis of $(1,2)$ forms
provided by \Ka\ covariant derivatives from its complex conjugate
$\widehat{\Omega}^{(0,3)}$.
This expression (\ref{WZTG}) is symplectic covariant.
Indeed, due to the special \Ka\ identity
\begin{equation}\label{identTG}
 -{\rm i}\XL^\ILambda T^+ +2\bar f_{\bar\imath}^\ILambda
 G^{+\,\bar\imath}= \mathcal{F}^{+\,\ILambda}\ ,
\end{equation}
it can be written as
\begin{equation}
S_{3 ,\rm WZ}=\prefactor\int_{M_5} \pmatrix{\widehat{\Omega}^{(3,0)}&
 \widehat{\Omega}_{\ib}
^{(1,2)}}
Y^{-1}{\cal F}^{I+}+ {\rm c.c.}\ ,
\end{equation}
where   $Y$ is the invertible $(n+1)\times(n+1)$ matrix
\begin{equation}
Y=\pmatrix{\XL^\ILambda & \bar f_{\bar\imath }^\ILambda}\ .
\end{equation}
The integral is performed over $M_5$ which is $M_2\times{\cal C}^3$,
the former being the manifold whose boundary is the worldline $M_1$
as in (\ref{0act}).
We perform the integral over the supersymmetric cycle ${\cal C}^3$,
using (\ref{periodi}), which can be written elegantly as
\begin{eqnarray}
\int_{{\cal C}^3} \pmatrix{\widehat{\Omega}^{(3,0)}& \widehat{\Omega}_{\bar\imath}^{(1,2)} }
&=&\sqrt{8\pi }\kk\pmatrix{p^\ILambda & q_\JSigma}\pmatrix{0&1\cr -1&0}
\pmatrix{ \XL^\JSigma &\bar f_{\bar\imath}^\JSigma\cr
\FM_\ILambda &\bar h_{\bar\imath \ILambda}}\nonumber\\ &=&\sqrt{8\pi }\kk
\pmatrix{p^\ILambda & q_\JSigma}\pmatrix{0&1\cr -1&0}\pmatrix{Y\cr {\cal N}Y}\ .
\end{eqnarray}
This allows us to write the WZ term as
\begin{eqnarray}
S_{3 ,\rm WZ}&=&\prefactor \,\sqrt{8 \pi}\kk \,
\int_{M_2}\pmatrix{p^\ILambda & q_\JSigma}\pmatrix{0&1\cr -1&0}
\pmatrix{Y\cr {\cal
N}Y} Y^{-1}\mathcal{F}^{\ILambda+}+ {\rm c.c.} \nonumber\\ &=&
\prefactor \,\sqrt{8 \pi}\kk\int_{ M_2}(-q_\ILambda +p^\JSigma{\cal
N}_{\JSigma\ILambda})\mathcal{F}^{\ILambda+}+ {\rm c.c.}\nonumber\\
&=&\prefactor\,\sqrt{8 \pi}\kk \int_{M_2}
(p^\JSigma \mathcal{G}_\JSigma-q_\ILambda \mathcal{F}^\ILambda )\ .
\end{eqnarray}
Using (\ref{holmag}) and (\ref{Diracqu}) one now establishes  that
$b(\sqrt{8\pi }\kk)4\pi \kk/(8\pi \kk^2)=\sqrt{2\pi }$, as we demanded
before, thus  $b=1$. With $\mu_3=\sqrt{2\pi }$ we find
the same result as in (\ref{S2ndorder}).
Putting together the Wess--Zumino and the kinetic term we have shown
that the wrapping of the \three brane action over a supersymmetric CY cycle
leads to the 0-brane action  derived previously, (\ref{S2ndorder}).
Let us note that in \cite{Iengo} it was shown that by wrapping a
D3-brane on a cycle of a $T^6/Z_3$ orbifold, one obtains a
Reissner--Nordstr\"om black hole solution of $\mathcal{N}=2$
supergravity.
\section{Generalization to higher ${\cal N}$ supergravity,
in particular ${\cal N}=8$}
In this section we discuss how our result  for the $0$-brane action coupled to $\mathcal{N}=2$
supergravity  can be extended to the case where the $0$-brane moves in a higher
extended supergravity background. In particular, we focus on the case ${\cal
N}=8$, for which the BPS black holes have been constructed and
classified in \cite{bholIt1},\cite{bholIt2},\cite{bholIt3}.
\par
As we show below the structure of the $\mathcal{N}=8$ 0-brane is very similar to
that of the  $\mathcal{N}=2$ brane and it is almost its straightforward
generalization.
Yet it is clear that for $\mathcal{N}>2$, there is the possibility
of having BPS configurations that preserve different fractions of
supersymmetry. In particular, the BPS black holes for the case
$\mathcal{N}=8$ fall into three different classes,
depending on the fraction of supersymmetry
\begin{equation}
\label{mb1}
{\nu\over 8}~, \hskip 0.3cm \nu=1,2,4~,
\end{equation}
that they preserve.
Similarly, we will discover that there are three classes of $\mathcal{N}=8$ $0$-brane actions where the worldline
$\kappa$-supersymmetry has, respectively, $1/2$, $1/4$ or $1/8$
of the $32= 8\times 4$  components possessed by $\mathcal{N}=8$
spacetime supersymmetry. Indeed, as it was already the case in
$\mathcal{N}=2$ supergravity, the appropriate projection
satisfied by the $\kappa$-supersymmetry parameter coincides
with the projection equation satisfied by the black hole
BPS Killing spinor.
It follows from these introductory remarks that, in order to
discuss the $\mathcal{N}=8$ $0$-brane actions we have to recollect some
results and properties of the corresponding black hole solutions.
\par
\begin{table}[hb]
\caption{The three classes of ${\cal N}=8$ BPS black holes}
\begin{center}
\begin{tabular}{c c c c}
\hline\hline
$\nu$ & SUSY & Central Charge & $G_{\rm stab}\subset  {\rm SU}(8)$ \\
\hline
1 & 1/8 &  $Z_1(\infty)\ne Z_2(\infty)\ne Z_3(\infty)\ne Z_4(\infty)$ &
${\rm USp}(2)^4$ \\
2 & 1/4 &  $Z_1(\infty)=Z_2(\infty)\ne Z_3(\infty)=Z_4(\infty)$ &
${\rm USp}(4)\times {\rm USp}(4)$\\
4 & 1/2 &  $Z_1(\infty)=Z_2(\infty)=Z_3(\infty)=Z_4(\infty)$ & USp$(8)$ \\
\hline\hline
\end{tabular}
\end{center}
\label{tbl:stabil}
\end{table}
As discussed
extensively in \cite{bholIt2,RiccardoPietroReview},
the three cases $\nu=1,2,4$ are characterized by the structure of
the central charge at infinity or, more intrinsically, by
the covariance group of the corresponding Killing spinor equation.
In $\mathcal{N}=8$ supergravity the central charge  $Z_{AB}$ is an
antisymmetric field tensor transforming in the {\bf {28}} representation
of $SU(8)$. By means of $SU(8)$ local transformations it can
always be brought to normal form, namely skew diagonalized as
follows:
\begin{equation}\label{skewZ}
   Z_{AB} =
{\rm diag}(Z_1\varepsilon,Z_2\varepsilon,Z_3\varepsilon,Z_4\varepsilon)
\ ;\qquad \varepsilon =\pmatrix{0&1\cr -1&0}~,
\end{equation}
and the complex numbers $Z_i(x)$ ($i=1,\dots,4$) are the four skew eigenvalues.
The
structure of these  eigenvalues at spatial infinity ($r \to \infty$) characterizes
the three distinct orbits of $\mathcal{N}=8$ black holes.
If we consider the ${\cal N}=8$ supersymmetry algebra generated by these Killing
spinors, where we have put $Z_{AB}$ in block-diagonal form (\ref{skewZ}),
then the analysis of the BPS bound on a state of mass $M$
reduces to the usual ${\cal N}=2$ analysis for each block.
Thus we must have
\begin{equation}
\label{mb2}
M\geq |Z_1|\geq |Z_2|\geq |Z_3|\geq |Z_4|~,
\end{equation}
having assumed in (\ref{skewZ}) a specific ordering of the eigenvalues.
When all $Z_i$ are different, the BPS bound can be saturated
in the first block only: $M=\pm |Z_1|^2$. If this is the case, a
combination of the supersymmetry charges with indices in the first block
annihilates the state, and 1/8 of supersymmetry is preserved.
When $Z_1=Z_2$, the saturation of the
BPS bound occurs simultaneously in the first two blocks: $M=|Z_1|=|Z_2|$,
leading to 2 preserved combination of supercharges (1/4 of the total number),
and so on. When all the eigenvalues of $Z$ are equal,
BPS saturation corresponds to conservation of 1/2 of supersymmetry.
The result is shown in table~\ref{tbl:stabil}.
\par
The stability subgroup $G_{\rm stab} \subset SU(8)$ of the central
charge is related to, but not identical to, the covariance group of
the Killing spinor equation. By the definition of the BPS state the
supersymmetry transformations vanish in the black hole background
if they are taken along a special supersymmetry parameter
$\epsilon_{A}$, the Killing spinor, satisfying a
suitable projection equation, which is the higher-${\cal N}$
generalization of (\ref{quispi}).
\par
 As formulated in \cite{myneucha,bholIt2,bholIt3},
such a condition is the following:
\begin{equation}
\label{kilspinor}
  \gamma_0 \,\epsilon_{A} = \varpi_{AB}  \epsilon^{B}~,
  \hskip 0.3cm
  \gamma_0 \,\epsilon^{A} = \varpi^{AB}  \epsilon_{B}~,
\end{equation}
where, in block notation,
\begin{equation}
\label{mb3}
 \varpi_{AB}=\varpi^{AB} =
{\rm diag}\,(\IC_{2\nu\times 2\nu},{\bf 0}_{(8-2\nu)\times
(8-2\nu)})~.
\end{equation}
The real antisymmetric matrix $\IC_{2\nu\times 2\nu}$ satisfies
$\IC_{2\nu\times 2\nu}^2$ $=-\bfone_{2\nu \times2\nu}$.
Notice that (\ref{kilspinor}) implies that the
Killing spinors are projected to the upper $2\nu\times 2\nu$ block:
\begin{equation}
\label{mb4}
\epsilon_{A} =  h_A{}^B \, \epsilon_{B}~,
\hskip 0.3cm
\epsilon^{A} =  h^A{}_B \, \epsilon^{B}~,
\end{equation}
by the matrix
\begin{equation}
\label{mb5}
h_A{}^B\equiv -\varpi_{AC} \, \varpi^{CB}
={\rm diag}\,\left(\bfone_{2\nu \times2\nu},{\bf 0}_{(8-2\nu)\times (8-2\nu)}\right)~.
\end{equation}
The three choices $\nu = 1,2,4$ correspond to $1/8$, $1/4$ and
$1/2$ of preserved supersymmetry, respectively.
\par
The invariance group $G_{\rm inv}$ of the
Killing spinor equation~(\ref{kilspinor}) is $G_{\rm inv}={\rm USp}(2\nu)
 \times {\rm U}(8-2\nu)$,
i.e.\ the $SU(8)$ subgroup that leaves the matrix $\varpi_{AB}$ unchanged.
\par
The structure of the Killing spinor projection is related to the
structure of these eigenvalues $Z_i$ of the central charge. Indeed, by direct
construction of the BPS black hole solutions  three distinct
possibilities, listed in table~\ref{tbl:stabil}, have been found.
\par
Comparing with our previous discussion we
see that the two groups $G_{\rm inv}$ and $G_{\rm stab}$ admit the common
factor $USp(2\nu)$. However, $G_{\rm inv}$ is bigger, as for its
determination only the blocks of $Z$ containing  the highest eigenvalue(s)
matter.
\par
The analogy between the ${\cal N}=8$ equation~(\ref{kilspinor})
and its ${\cal N}=2$ counterpart (\ref{quispi}) is complete only
if, in the latter case,  we get rid of the phase $Z/|Z|$ by
imposing reality of the central charge $Z=|Z|$. In the  ${\cal
N}=2$ theory  this is a $U(1)$ gauge choice   that can  always be
reached by means of suitable K\"ahler transformations. However, as
our previous experience teaches, if we want a $U(1)$ gauge
covariant form of the $\kappa$-supersymmetric action we had
better keep the phase $Z/|Z|$ in place and write the projection
on the $\kappa$-symmetry parameter as  given in (\ref{proje}).
So, although the black hole solutions of ${\cal N}=8$
supergravity have been derived starting from
equations~(\ref{kilspinor}), the previous observations suggest that such a
construction is based on a $SU(8)$ gauge-choice that we had
better undo in order to obtain an $SU(8)$ covariant form of the
$0$-brane $\kappa$-supersymmetric actions. Indeed, the role of
$U(1)$ is now played by the group $SU(8)$ which is both the
automorphism group of the ${\cal N}=8$ supersymmetry algebra and
the isotropy subgroup of the homogeneous scalar manifold.
\par
To discuss the $SU(8)$ covariantization of the Killing spinor
equation~(\ref{kilspinor}) we have to review the basic ingredients
of the ${\cal N}=8$ theory and fix our conventions.
\subsection{Scalar fields, $E_{7(7)}$ structure and 
the central charge} 
The theory possesses 70 scalar fields $\phi^I$ that span the non-compact coset
manifold $E_{7(7)}/SU(8)$. Using standard notations (see, for
example,
\cite{RiccardoPietroReview}) these fields are introduced through
the coset representative $\IL(\phi)$, that is an $E_{7(7)}$ matrix in
the fundamental ${\bf 56}$ representation. We have
\begin{equation}
\label{Lcoset}
\IL={1 \over\sqrt{2}}\left(\begin{array}{c|c}
                             f+\mbox{\rm i} h & \bar{f}+\mbox{\rm i}\bar{h}\\
                             \hline\\
                             f-\mbox{\rm i} h & \bar{f}-\mbox{\rm i}\bar{h}
                             \end{array}
                     \right)\ ,
\end{equation}
where the $28\times28$ submatrices $\left(h,f\right)$ are labeled
by antisymmetric pairs $(\Lambda,\Sigma)$ and $(A,B)$, with
$\Lambda,\Sigma=1,\dots,8$ and $A,B=1,\dots,8$, the first
pair transforming under $E_{7\left(7\right)}$ and the second one under
$SU(8)$:
\begin{equation}
\left(h,f\right)=\left(h_{\Lambda\Sigma|AB},
f^{\Lambda\Sigma}{}_{AB}\right)\ .
\end{equation}
As expected from general arguments we have $\IL\in USp\left(28,28\right)$.
Indeed, the theory contains $28$ gauge bosons so that we have $28$
electric field strengths and $28$ magnetic ones which have to
transform into one another through elements of the $USp(28,28)$
group.
The vielbein $P_{ABCD}$ and the $SU(8)$ connection
$\Omega_{A}{}^{B}$ of $E_{\left(7\right)7}/ SU\left(8\right)$ are
computed from the left-invariant $1$-form $\IL^{-1}d\IL$:
\begin{equation}
\label{L-1dL}
\IL^{-1}d\IL=\left(\begin{array}{c|c}
                             \delta^{[A}{}_{[C}\Omega^{B]}{}_{D]} & \bar{P}^{ABCD}\\
                             \hline\\
                             P_{ABCD} & \delta_{[A}^{}{[C}\bar{\Omega}_{B]}{}^{D]}\end{array}
                     \right)\ ,
\end{equation}
where $P_{ABCD}\equiv
P_{ABCD, i}d\Phi^{i}$, with $\left(i=1,\dots,70\right)$, is
completely antisymmetric and satisfies the reality condition
\begin{equation}
\label{realcondP}
P_{ABCD}={1 \over 24}\epsilon_{ABCDEFGH}\bar P^{EFGH}\ .
\end{equation}
The bosonic Lagrangian of $\mathcal{N}=8$ supergravity is \cite{cre}
\begin{eqnarray}
{\cal L}&=&\int\sqrt{-g}\, d^4x\left(-\ft12 R+\ft14\Im{\cal
N}_{\Lambda\Sigma|\Gamma\Delta}F_{\mu\nu}{}^{\Lambda\Sigma}F^{\Gamma\Delta|\mu\nu}-
\ft16P_{ABCD,i}\bar{P}^{ABCD}_{j}\partial_{\mu}
\Phi^{i}\partial^{\mu}\Phi^{j}+\right.\nonumber\\
&+&\left.\ft18\Re{\cal N}_{\Lambda\Sigma|\Gamma\Delta}
{\epsilon^{\mu\nu\rho\sigma}\over\sqrt{-g}}
F_{\mu\nu}{}^{\Lambda\Sigma}F^{\Gamma\Delta}{}_{\rho\sigma}\right)
\label{lN=8}\ ,
\end{eqnarray}
where  the gauge kinetic matrix ${\cal N}_{\Lambda\Sigma|\Gamma\Delta}$ is
defined  by ${\cal N}=hf^{-1}$, i.e.\ explicitly by
\begin{equation}
\label{defN}
{\cal N}_{\Lambda\Sigma|\Gamma\Delta}=
h_{\Lambda\Sigma|AB}\,(f^{-1})^{AB}{}_{\Gamma\Delta}\ .
\end{equation}
The same matrix relates the (anti)self-dual electric and magnetic
$2$-form field strengths:
\begin{equation}
\label{defG}
G^{-}_{\Lambda\Sigma}=
\bar{{\cal N}}_{\Lambda\Sigma|\Gamma\Delta}F^{-~\Gamma\Delta}~,
\hskip 0.3cm
G^{+}_{\Lambda\Sigma}={\cal
N}_{\Lambda\Sigma|\Gamma\Delta}F^{+~\Gamma\Delta}~,
\end{equation}
where the dual field strengths $G^{\pm}_{\Lambda\Sigma}$,  are,
as before, defined
by $G^{\pm}_{\Lambda\Sigma}={i \over 2}{\delta{\cal L} \over
\delta F^{\pm~\Lambda\Sigma}}$.
Note that the $56$-dimensional (anti)self-dual vector $\left(F^{\pm~\Lambda\Sigma},
G^{\pm}_{\Lambda\Sigma}\right)$
transforms covariantly under $Sp\left(56,\IR\right)$.
The matrix transforming the coset representative $\IL$ from the
$USp\left(28,28\right)$ basis, (\ref{Lcoset}), to the real
$Sp\left(56,\IR\right)$ basis is the Cayley matrix:
\begin{equation}
\IL_{\rm USp}={\cal C}\IL_{Sp}{\cal C}^{-1}\ ;\qquad {\cal C}=\pmatrix{ \bfone &
\mbox{\rm i} \bfone
\cr \bfone &-\mbox{\rm i}\bfone \cr}\ .
\label{caylone}
\end{equation}
Having established our definitions and notations, let us now write
the {\em dressed graviphoton} $2$-form,
  defined according to the obvious generalization of (\ref{G-def}):
\begin{equation}
T^{\left(-\right)}_{AB}=h_{\Lambda\Sigma AB}
\left(\Phi\right)F^{-\Lambda\Sigma}-f^{\Lambda\Sigma}{}_{AB}
\left(\Phi\right)G^-_{\Lambda\Sigma}\ .
\label{tab}
\end{equation}
In a way similar  to the ${\cal N}=2$
case   we also have the identities:
\begin{eqnarray}
&&T^+_{AB}=0\rightarrow T^-_{AB}=T_{AB}\ ;\qquad
\bar{T}^-_{AB}=0\rightarrow \bar{T}^{+\,AB}=\bar{T}^{AB}\ .
\end{eqnarray}
Thus we can define the central charge:
\begin{equation}
Z_{AB}=\frac{1}{\pi }\int_{S^2}{T_{AB}}=h_{\Lambda\Sigma|AB}
p^{\Lambda\Sigma}-f^{\Lambda\Sigma}{}_{AB}q_{\Lambda\Sigma}\ ,
\label{zab2}
\end{equation}
which, as we already anticipated, is an antisymmetric tensor transforming in the
${\bf {28}}$ irreducible representation of $SU(8)$.
In (\ref{zab2}) the integral of the 2-form $T_{AB}$ is evaluated on any static
$2$-sphere  and the quantized charges ($p^{\Lambda\Sigma},~q_{\Lambda\Sigma}$)
are defined, in analogy with  (\ref{holmag}) (choosing in this section
the normalization with $\kk=1/4$) by
\begin{equation}
p^{\Lambda\Sigma}=\frac{1}{\pi}\int_{S^2}{F^{\Lambda\Sigma}}~,
\hskip 0.3cm
q_{\Lambda\Sigma}=\frac{1}{\pi}\int_{S^2}G_{\Lambda\Sigma}\ .
\label{pqcharge}
\end{equation}
Given these preliminaries let us now review the  general form of
the $1/2$ and $1/4$ BPS black hole solutions as constructed in
\cite{bholIt2} and compute the corresponding field-dependent central charge.
This gives us the opportunity to see, through explicit formulae, how the
Killing spinor equation~(\ref{kilspinor}) can be translated in
terms of the central charge and in this way $SU(8)$ covariantized.
\subsection{The $1/2$ Black-hole and its central charge}
In the 1/2 supersymmetry-preserving case ($\nu=4$), the Killing spinor
projection (\ref{kilspinor}) involves the matrix
\begin{equation}\label{unmezom}
  \varpi=\IC_{8\times 8}\ ,
\end{equation}
so that the projector $h_{A}{}^{B}$ is simply $\delta_A{}^{B}$.
Recalling the results  of \cite{bholIt2},
let us write the BPS black hole solution admitting the gauge-fixed
Killing spinor (\ref{kilspinor}), that contains a new parameter $q$.
The metric is
\begin{equation}\label{12met}
  ds^2 = -\left [ H(x)\right ]^{-1/2} \, dt^2 + \left [ H(x)
  \right ]^{1/2} d\vec{x}^2\ .
\end{equation}
The electric and magnetic field strengths are, respectively,
\begin{eqnarray}
  {\cal F}^{\Lambda\Sigma} &=& - \, q \, \IC^{\Lambda\Sigma} \, \left [ H(x) \right ]^{-2} \,
    \frac{1}{4r^3} dt \, \wedge \, \vec{x}\cdot
    d\vec{x}~,\nonumber\\
  {\cal G}_{\Lambda\Sigma} &= &- \,  q \,  \IC_{\Lambda\Sigma} \, \left [ H(x) \right ]^{1/2} \,
    \frac{x^i}{8r^3}\, dx^j \, \wedge \, dx^k \,
    \epsilon_{ijk}\label{12G}\ .
\end{eqnarray}
The scalar fields are described through the coset representative
\begin{eqnarray}
  f^{\Lambda\Sigma}_{AB} & =& \frac{1}{8\sqrt{2}} \IC^{\Lambda\Sigma} \, \IC_{AB}
   \, \left [ H(x) \right ]^{-3/4}~, \nonumber\\
   h_{\Lambda\Sigma \vert AB} & =&  \,
   -\frac{\ii}{8\sqrt{2}} \IC_{\Lambda\Sigma} \, \IC_{AB}
   \, \left [ H(x) \right ]^{ 3/4}\ , \label{12hmat}
\end{eqnarray}
where $ H(x) $ denotes a harmonic function in the 3-dimensional
transverse space with boundary condition $H(\infty)=1$. Typically
one has
\begin{equation}\label{harmonio}
    H(x)   = 1+ \frac{q}{r}\ ,
\end{equation}
where $q$ is the same parameter appearing in (\ref{12G}).
\par
Inserting (\ref{12G}) in the integrals
(\ref{pqcharge}) we obtain the values of the electric and magnetic
charges for this solution:
\begin{equation}\label{pq12}
  p^{\Lambda\Sigma} = 0 \quad ; \qquad q_{\Lambda\Sigma} = - \, q \,
  \IC_{\Lambda\Sigma}\ ,
\end{equation}
and using (\ref{pq12}) and (\ref{12hmat}) into
(\ref{zab2}) we obtain the central charge:
\begin{equation}\label{zab12}
  Z_{AB} = \frac{q}{\sqrt{2}} \, \left [ H(x) \right ]^{-3/4} \,
  \IC_{AB}\ .
\end{equation}
{}From the explicit form (\ref{zab12}) we see that the central
charge at spatial infinity approaches an antisymmetric matrix with
four coinciding skew eigenvalues:
\begin{equation}\label{zabinf12}
  Z_{AB} \, \stackrel{ r \to \infty}{\longrightarrow} \,\frac{q}{\sqrt{2}}
  \,\IC_{AB}\ ,
\end{equation}
as expected from table~\ref{tbl:stabil}. On the other
hand, near the horizon $r \to 0$ the central charge goes to zero as
$r^{3/4}$. This confirms what we also expect on general grounds,
namely that the entropy of the black hole, proportional to the horizon
value of $(Z_{AB} \, {\bar Z}^{AB} )^{1/2}$ is  zero in the $1/2$
SUSY case.
\par
As  we stressed at the beginning of this section, by writing the Killing
spinor equation in the form (\ref{kilspinor}) we have worked in
a fixed $SU(8)$ gauge. Yet it is now quite easy to relax this
gauge choice by performing an arbitrary, local $SU(8)$ transformation on
the  expression we have obtained for the central charge. Let $U(x)
\in SU(8)$ be such a gauge transformation. For the BPS $1/2$ SUSY
preserving black hole we obtain
\begin{eqnarray}\label{UZUt12}
  Z_{AB}(x)& =& \lambda (x)  \left [ U(x) \, \IC \, U^T(x) \right
  ]_{AB}\ , \end{eqnarray}
  where
  \begin{eqnarray}\label{unlambda}
  \lambda(x) & = & \bar \lambda(x) \equiv \, \frac{q}{\sqrt{2}} \,
  \left [ H(x) \right ]^{ -3/4}
\end{eqnarray}
is a unique {\em  real} skew eigenvalue characterizing the central
charge at any point in spacetime. Since by definition of
the $USp(8)$ group the matrix $\IC$ is invariant against such
transformations, it follows that in (\ref{UZUt12}) we have introduced
$63-36=27$ new arbitrary functions  parametrizing the coset
manifold $SU(8)/USp(8)$. They  are the ${\cal N}=8$ analogue of the single
phase $Z/\vert Z \vert $ appearing in the ${\cal N}=2$ theory.
Yet, what matters and is an intrinsic property of the $1/2$ background
is that: {\sl in any spacetime point the four skew eigenvalues
of the central charge $Z_{AB}$ are real and coincide.}
\par
{}From this property follows an identity which will be crucial in proving
the $\kappa$-supersymmetry of the $0$-brane action.
Indeed, using the expression (\ref{UZUt12}) for the central charge, we have
\begin{eqnarray}
\ft{1}{24} \, \epsilon^{ABCDEFGH} \, Z_{EF} \, Z_{GH}
&=&\ft{1}{24}\,\lambda(x) \,\epsilon^{A B C D E F G H}\,
{  U}_{E}^{\phantom{E}E^\prime}\,{  U}_{F}^{\phantom{F}F^\prime}
\,{  U}_{G}^{\phantom{G}G^\prime}\,{  U}_{H}^{\phantom{H}H^\prime}
\, C_{ E^\prime F^\prime} \, C_{G^\prime H^\prime} \nonumber\\
&=&\ft{1}{24}\,\lambda(x)\,
{\bar U}^{A}_{\phantom{A}A^\prime}\,{\bar U}^{B}_{\phantom{B}B^\prime}
\,{\bar U}^{C}_{\phantom{C}C^\prime}\,{\bar U}^{D}_{\phantom{D}D^\prime}
\, \epsilon^{A^\prime B^\prime C^\prime D^\prime E^\prime F^\prime G^\prime H^\prime}\,
C_{ E^\prime F^\prime} \, C_{G^\prime H^\prime} \nonumber\\
&=&\,\lambda(x)\,{\bar U}^{A}_{\phantom{A}A^\prime}\,{\bar U}^{B}_{\phantom{B}B^\prime}
\,{\bar U}^{C}_{\phantom{C}C^\prime}\,{\bar U}^{D}_{\phantom{D}D^\prime}
\, C^{[ A^\prime B^\prime} \, C^{C^\prime D^\prime]} \nonumber\\
&=&{\bar Z}^{[AB} \, {\bar Z}^{CD]}\ .\label{idezz}
\end{eqnarray}
Note that the last equality follows precisely from the reality of the skew
eigenvalue $\lambda(x)$.
\par
Let us now turn our attention to the $1/4$ susy preserving
black holes.
\subsection{The $1/4$ black hole and its central charge}
In the $1/4$ case ($\nu=2$) we have:
\begin{equation}\label{unquaom}
  \varpi_{AB}=-\varpi_{BA}={\rm diag}\,(\IC_{4 \times 4},{\bf 0}_{4 \times 4})\ ,
\end{equation}
and following \cite{bholIt2} it is convenient to introduce the further notations:
\begin{eqnarray}\label{omOmtau}
  \Omega_{AB}= -\Omega_{BA} & = &
  {\rm diag}\,({\bf 0}_{4 \times 4},\IC_{4 \times 4})~,
  \nonumber\\
  \tau^\pm_{AB} \equiv \ft{1}{2}\, \left ( \varpi_{AB} \pm
  \Omega_{AB} \right ) & = &
  \ft12\,{\rm diag}\,(\IC_{4 \times 4},\pm\IC_{4 \times 4})~.
\end{eqnarray}
Then  the BPS black hole solution admitting the gauge-fixed
Killing spinor (\ref{kilspinor}) can be written as follows.
The metric is
\begin{equation}\label{14met}
  ds^2 =- \left [ H_1(x) \,H_2(x)\right ]^{-1/2} \, dt^2 +
  \left [ H_1(x) \,H_2(x) \right ]^{1/2} d\vec{x}^2\ .
\end{equation}
The electric and magnetic field  strength are
\begin{eqnarray}
  {\cal F}^{\Lambda\Sigma} &=& - \,\frac{1}{2\sqrt{2}} \,
   \left (\left [ H_1(x) \right]^{-2} \,q_1 \, \tau^+_{\Lambda\Sigma} \, + \,
\left [ H_2(x) \right]^{-2} \,q_2 \, \tau^-_{\Lambda\Sigma} \right ) \,
    \frac{1}{4r^3} dt \, \wedge \, \vec{x}\cdot
    d\vec {x}~,\nonumber\\
  {\cal G}_{\Lambda\Sigma} &=& - \,\frac{1}{8\sqrt{2}} \,
   \left (q_1 \,\left [ \frac{H_2(x)}{H_1(x)} \right]^{2} \, \tau^+_{\Lambda\Sigma} \, + \,
q_2 \,\left [ \frac{H_1(x)}{H_2(x)} \right]^{2} \, \tau^-_{\Lambda\Sigma} \, \right ) \,
    \frac{x^i}{8r^3}\, dx^j \, \wedge \, dx^k \,
    \epsilon_{ijk}\label{14G}\ .
\end{eqnarray}
The coset representatives describing the scalar fields are
\begin{eqnarray}
   f^{\Lambda\Sigma}_{AB} & =& \frac{1}{2\sqrt{2}} \left (
  \left[\frac{H_2(x)}{H^3_1(x)}\right ]^{1/4}  \,
  \tau^+_{\Lambda\Sigma} \, \tau^+_{AB} \, + \,
   \left[\frac{H_1(x)}{H^3_2(x)}\right ]^{1/4}  \,
  \tau^-_{\Lambda\Sigma} \, \tau^-_{AB} \, \right )~, \nonumber\\
  h_{\Lambda\Sigma\vert AB} & =& \frac{-\rm i}{2\sqrt{2}}\, \left (
  \left[\frac{H_1(x)}{H^3_2(x)}\right ]^{-1/4}  \,
  \tau^+_{\Lambda\Sigma} \, \tau^+_{AB} \, + \,
   \left[\frac{H_2(x)}{H^3_1(x)}\right ]^{-1/4} \,
  \tau^-_{\Lambda\Sigma} \, \tau^-_{AB} \, \right )~, \label{14hmat}
\end{eqnarray}
where $ H_{1,2}(x) $ are two harmonic functions in the 3-dimensional
transverse space with boundary condition $H(\infty)=1$:
\begin{equation}\label{armonie}
    H_1(x)   = 1+ \frac{q_1}{r} \quad ; \qquad  H_2(x)   = 1+ \frac{q_2}{r}
\end{equation}
where $q_{1,2}$ is the same parameter appearing in (\ref{14G}).
\par
Inserting (\ref{14G}) in the integrals
(\ref{pqcharge}) we obtain the values of the electric and magnetic
charges for this solution:
\begin{equation}\label{pq14}
  p^{\Lambda\Sigma} = 0 \quad ; \qquad q_{\Lambda\Sigma} =
   - \,\frac{1}{8\sqrt{2}} \, \left [ q_1 \, \tau^+_{\Lambda\Sigma} + q_2 \,
   \tau^-_{\Lambda\Sigma} \right ]\ ,
\end{equation}
and using (\ref{pq14}) and (\ref{14hmat}) in
(\ref{zab2}) we obtain the central charge:
\begin{equation}\label{zab14}
  Z_{AB} = -\frac{1}{16} \, \left(
  q_1 \,\left [\frac{H_2(x)}{H^3_1(x)} \right ]^{1/4} \, \tau^+_{AB}
  \, + \,
  q_2 \,\left [\frac{H_1(x)}{H^3_2(x)} \right ]^{1/4} \, \tau^-_{AB}
  \right )\ .
\end{equation}
{}From the explicit form (\ref{zab14}) we see that  at spatial
infinity the central charge approaches the antisymmetric matrix
\begin{equation}\label{zabinf14}
  Z_{AB}^\infty =-\ft{1}{16} \,
  \left( q_1 \, \tau^+_{AB}\, + \,q_2 \, \tau^-_{AB} \right ) = -\ft{1}{16}
  \,{\rm diag}\,\left((q_1+q_2) \, \IC_{4 \times 4}, (q_1-q_2) \,  \IC_{4 \times 4} \
  \right)\ ,
\end{equation}
whose four skew eigenvalues are real and coincide in pairs as expected
from table~\ref{tbl:stabil}. On the other
hand, near the horizon $r \to 0$ the central charge goes to zero as
$r^{1/2}$.  Also in this case the entropy is zero.
\par
Once again, we can remove the $SU(8)$ gauge fixing utilized in
deriving the $1/4$ BPS solution by writing the analogue of
 (\ref{UZUt12}), namely
\begin{equation}\label{UZUt14}
  Z_{AB}= \left [ U(x)  \left( \lambda_1(x) \, \tau^+_{AB} \, + \,
   \lambda_2(x) \, \tau^-_{AB} \right ) \, U^T(x) \right ]_{AB}~,
\end{equation}
where
\begin{equation}
  \lambda_1(x)=-\frac{q_1}{16} \,\left [\frac{H_2(x)}{H^3_1(x)} \right ]^{1/4}~,
  \hskip 0.3cm
   \lambda_2(x)=-\frac{q_2}{16} \,\left [\frac{H_1(x)}{H^3_2(x)} \right
   ]^{1/4}\ ,
\end{equation}
and $U(x)$ is an arbitrary $SU(8)$ gauge transformation.
\subsection{The $\kappa$-supersymmetry projection and the $0$-brane action for the
cases of $1/2$, $1/4$ and $1/8$ BPS backgrounds}
 Having clarified
these preliminaries we can now proceed to write the appropriate
$SU(8)$ covariant form of the projection on the
$\kappa$-supersymmetry parameter $\kappa_A , \kappa^A$ and the
worldline action that is invariant against such $\kappa$
transformations.
\par
Recalling equation~(\ref{mb5}) we introduce an $ 8 \times
8$ projection matrix:
\begin{equation}\label{8proj}
  {\cal P}^{(2\nu)B}_A(x) \equiv \left [ U^\dagger (x) \, h^{(2\nu)} \, U(x) \right ]_A
  {}^B~,
\end{equation}
where $U(x) \in SU(8)$ is the point-dependent $SU(8)$ gauge
transformation that skew-diagonalizes the central charge.   Such
a transformation was already introduced in
(\ref{UZUt12}) and (\ref{UZUt14}). The parameter $\nu$ takes the
values $\nu=4,2,1$ and distinguishes among the three cases of
$\nu/8$ preserved supersymmetries. According to the conventions
introduced in equation~(\ref{kilspinor}), $2\nu$ is the rank of the
projector ${\cal P}^{(2\nu)A}_B(x)$.  Next we introduce the
projected central charge tensor:
\begin{eqnarray}\label{prozeta}
  \widehat{Z}_{AB}={\cal P}^{(2\nu)C}_A(x)\,{\cal
  P}^{(2\nu)D}_B(x)\, Z_{CD}\ .
\end{eqnarray}
Using such a notation the projection on either the Killing
spinor of the BPS background (\ref{kilspinor}) or the
$\kappa$-supersymmetry parameter can be rewritten in a manifestly
$SU(8)$ covariant fashion as follows:
\begin{eqnarray}\label{su8ksym}
\Pi^a\gamma_a\kappa_A & = &
\sqrt{2\nu}\, \left(\widehat{Z}_{CD} \,
\widehat{\overline{Z}}^{CD} \right)^{-1/2} \, \widehat{Z}_{AB} \, \kappa^B~,
\nonumber\\
\Pi^a\gamma_a\kappa^A & = &
\sqrt{2\nu}\, \left(\widehat{Z }_{CD} \,
\widehat{\overline{Z}}^{CD} \right)^{-1/2} \,\widehat{\overline{Z}}^{AB} \,
\kappa_B~.
\end{eqnarray}
The generalization of the condition (\ref{mb4}) is
\begin{equation}
\label{mb6}
  \kappa_A = {\cal P}_A{}^{B} \,  \kappa_B~,\hskip 0.3cm
  \kappa^A = \bar{\cal P}^A{}_{B} \,\kappa^B\ .
\end{equation}
The $\kappa$-supersymmetric action of a $0$-brane moving in the
background of a $\nu/8$ supersymmetry preserving  ${\cal N}=8$
on-shell configuration can now be written. It is an almost direct
generalization of (\ref{0act}) and reads as follows:
\begin{eqnarray}
\frac{1}{\pi }S_{(2\nu)}^{\mathcal{N}=8}&=& \int_{M_1} \, - \alpha_{2\nu}\,
\left ( \widehat{Z}_{AB}\, \widehat{{\bar Z}}^{AB} \right )^{1/2}
\, \left [- \left (\Pi^a \, V^b + \ft{1}{2} \, \Pi^a \, \Pi^b \,
e \right) \, \eta_{ab} +\ft{1}{2} \, e \right] \nonumber\\ &&+
\int_{M_2} \, \left( p^{\Lambda\Sigma} \, {\cal
G}_{\Lambda\Sigma} - q_{\Gamma\Delta} \, {\cal F}^{\Gamma\Delta}
\right)\ , \label{0actn8}
\end{eqnarray}
the parameter $\alpha_{2\nu}$ being fixed by $\kappa$-invariance:
\begin{equation}
\alpha_{2\nu}=\sqrt{\frac{2}{\nu}}\ .\label{alpcof}
\end{equation}
The proof of $\kappa$-supersymmetry can be done along the same
lines as in the ${\cal N}=2$ case; there are, however, a few
subtleties that have to be taken into account, and that are
different in the three cases $\nu=4,2,1$.
{}First of all we need to recall
the supersymmetry transformation rules of ${\cal N}=8$
supergravity. They read as follows (see \cite{cre} and for current
notations \cite{bholIt1,bholIt2,bholIt3} and
\cite{RiccardoPietroReview}):
\begin{eqnarray}
\delta\chi_{ABC}&=&~4 ~P_{ABCD|i}\partial_{\mu}\Phi^i\gamma^{\mu}
\epsilon^D+3T^{(-)}_{[AB|\rho\sigma}\gamma^{\rho\sigma}
\epsilon_{C]} \label{chidelta}~,\\
\delta\psi_{A\mu}&=&\nabla_{\mu}\epsilon_A~
~+\frac{\rm i}{4}\, T^{(-)}_{AB|\rho\sigma}\gamma^{\rho\sigma}\gamma_{\mu}
\epsilon^{B}~, \label{psidelta} \\
\delta A^{\Lambda\Sigma}_\mu & = & 2\,  {\bar f}^{\Lambda\Sigma\vert AB}
\bar \psi_{A\mu} \epsilon_B
\epsilon^{AB}\,+\, 2\, f^{\Lambda\Sigma}_{AB} \bar\psi^A_{\mu}\epsilon^B \epsilon
_{AB}\label{adelta}\\
& &+\,\ft{1}{4}\, \left ( {\bar f}^{\Lambda \vert AB} \,
 {\bar \chi}_{ABC} \, \gamma_a \, \epsilon^C + f^\Lambda_{AB} {\bar \chi}^{ABC}
 \, \gamma_a \, \epsilon_C \right ) \, V^a_\mu~, \nonumber\\
  P^{ABCD}_i \delta \Phi^i & = & {\bar \chi}^{[ABC} \, \epsilon^{D]}  \, + \,
  \ft{1}{24} \, \varepsilon^{ABCDPQRS} \, {\bar \chi}_{PQR} \, \epsilon_S
  ~,\label{pidelta}\\
  P_{ABCD\vert i} \delta \Phi^i & = & {\bar \chi}_{[ABC} \, \epsilon_{D]}  \, + \,
  \ft{1}{24} \, \varepsilon_{ABCDPQRS} \, {\bar \chi}^{PQR} \, \epsilon^S
  ~,\label{bpidelta}\\
  \delta V^a_\mu &=& {\bar \epsilon}^A \, \gamma^a \, \psi_{A\mu} +
 {\bar \epsilon}_A \, \gamma^a \, \psi^{A\mu}\ . \label{vdelta}
\end{eqnarray}
As before the $\kappa$-supersymmetry is an ordinary supersymmetry
transformation of the background fields with a supersymmetry
parameter satisfying a suitable projection, given
in this case by equation~(\ref{su8ksym}). The variation of
the $0$-brane action is done as in (\ref{psivar}),
(\ref{lamka}) and (\ref{lavario}) and we have to check the separate
cancellation of the gravitino and dilatino terms.
\par
Just as before let us begin with the gravitino terms. At this
level we obtain the variation:
\begin{eqnarray}\label{psivar8}
 \frac{1}{\pi }   \delta S_{(2\nu)}^{\mathcal{N}=8}& = &
    \int_{M_1}   \alpha_{2\nu} \,
    \left ( \widehat{Z}_{AB}\, \widehat{{\bar Z}}^{AB} \right )^{1/2}
    \left [ \, \overline{\psi}_A \,\gamma^a \, \kappa^A
    + \, \overline{\psi}^A \,\gamma^a \, \kappa_A \right ] \,  \Pi_a\nonumber\\
    \null & \null & +    2 \, p^{\Lambda\Sigma} \int_{M_1}
    \left(\overline{h}_{\Lambda\Sigma}^{\phantom{\Lambda\Sigma}AB} \,
    \overline{\psi}_A \kappa_B +
     h_{ \Lambda\Sigma\vert AB} \,
    \overline{\psi}^A \kappa^B \right )\nonumber\\
    \null&\null  & -    2 \, q_{\Lambda\Sigma} \int_{M_1}
    \left(\overline{f}^{ \Lambda\Sigma\vert AB} \,
    \overline{\psi}_A \kappa_B +
     f^{\Lambda\Sigma}_{\phantom{\Lambda\Sigma} AB} \,
    \overline{\psi}^A \kappa^B \right ) \ .
\end{eqnarray}
The second and third lines in equation~(\ref{psivar8}) reconstruct
the definition of the central charge tensors $Z_{AB}, {\bar
Z}^{AB}$. Yet, because of (\ref{mb6}),
we can safely replace the central charges
with their hatted counterparts:
\begin{equation}\label{hatzgood}
  \begin{array}{ccl}
   \null & \null & 2 \, p^{\Lambda\Sigma} \,
    \left(\overline{h}_{\Lambda\Sigma}^{\phantom{\Lambda\Sigma}AB} \,
    \overline{\psi}_A \kappa_B +
     h_{ \Lambda\Sigma\vert AB} \,
    \overline{\psi}^A \kappa^B \right )  -
   2 \, q_{\Lambda\Sigma} \,
    \left(\overline{f}^{ \Lambda\Sigma\vert AB} \,
    \overline{\psi}_A \kappa_B +
     f^{\Lambda\Sigma}_{\phantom{\Lambda\Sigma} AB} \,
    \overline{\psi}^A \kappa^B \right )\\
    \null  &= &
    2 \widehat{\overline{ Z} }^{AB}\, \overline{\psi}_A \kappa_B\,
    + \, 2 \widehat{{  Z}}_{AB}\, \overline{\psi}^A \kappa^B\ . \\
  \end{array}
\end{equation}
In the first line of   (\ref{psivar8}) we can reduce the
terms involving the gamma matrix $\gamma^a$ by use of the
projection (\ref{su8ksym}) and we obtain
cancellation of the gravitino terms if the parameter
$\alpha_{2\nu}$ satisfies the condition (\ref{alpcof}). Just as
in the ${\cal N}=2$ case, the non-trivial check is to verify that the
same coefficient also guarantees  the cancellation of the dilatino
terms. This indeed happens in a different
subtle way for the three values $\nu =4,2,1$.
\par
At the level of the dilatino terms the $\kappa$-supersymmetry
variation of the $0$-brane action reads as follows:
\begin{eqnarray}\label{dilvar}
\frac{1}{\pi} \delta S_{(2\nu)}^{\mathcal{N}=8} & = &
\int_{M_1}   - {\alpha_{2\nu}\over 2}
\left (\widehat{Z}_{CD}\widehat{{\bar Z}}^{CD} \right )^{-1/2}
\biggl( \widehat{\overline{{Z}}}^{CD}
     P_{CDRS,i} \delta_\kappa \Phi^i
     \overline{Z}^{RS}+\widehat{Z}_{CD}
     \overline{P}^{CDRS}_{ i} \delta_\kappa \Phi^i
      {Z}_{RS}\biggr)\nonumber\\
& + &\ft{1}{4}  \, p^{\Lambda\Sigma} \int_{M_1}
    \left(\overline{h}_{\Lambda\Sigma}^{\phantom{\Lambda\Sigma}PQ} \,
    \overline{\chi}_{PQC} \gamma_a \kappa^C +
     h_{ \Lambda\Sigma\vert PQ} \,
    \overline{\chi}^{PQC} \gamma_a \kappa_C\right ) \, V^a\nonumber\\
& - &  \ft{1}{4}   \, q_{\Lambda\Sigma} \int_{M_1}
    \left(\overline{f}^{ \Lambda\Sigma\vert PQ} \,
    \overline{\chi}_{PQC} \gamma_a \kappa^C +
     f^{\Lambda\Sigma}_{\phantom{\Lambda\Sigma} PQ} \,
     \overline{\chi}^{PQC} \gamma_a \kappa_C\right ) \, V^a \ .
\end{eqnarray}
In deriving the above equation we have used  the following
differential relation satisfied by the central charge
\cite{LauSerRiczent} and therefore by its projected version:
\begin{equation}\label{ergozhat}
 \begin{array}{ccccccc}
   \nabla  Z_{AB} & = &    P_{ABCD}\overline{ Z}^{CD}& \longrightarrow &
   \nabla \widehat{Z}_{AB} & = &  {\cal P}_A^{A^\prime}\,
    {\cal P}_B^{B^\prime} P_{A^\prime B^\prime CD}\,
    \overline{Z}^{CD}\ ,
 \end{array}
  \end{equation}
the symbol $\nabla$ denoting the $SU(8)$ covariant derivative.
\par
Substituting the explicit form of the scalar field supersymmetry
transformation (\ref{pidelta}) and utilizing the projection
(\ref{su8ksym}) we obtain:
\begin{eqnarray}\label{puffo}
\frac{1}{\pi } \delta S_{(2\nu)}^{\mathcal{N}=8}
& = & \int_{M_1}  - {\alpha_{2\nu}\over 2}
    \left(\widehat{Z}_{CD}\widehat{{\bar Z}}^{CD} \right )^{-1/2}
    \left(\widehat{\overline{{Z}}}^{CD}\overline{Z}^{RS}
    +\ft{1}{24}\widehat{Z}_{IJ}{Z}_{KL}\epsilon^{IJKLCDRS} \right)
    \overline{\chi}_{[CDR} \,\kappa_{S]}\nonumber\\
& + & {\sqrt{2\nu}\over 4}\,
    \left (\widehat{Z}_{CD}\widehat{{\bar Z}}^{CD} \right)^{-1/2}\,
    \overline{ Z }^{CD} \,
    \widehat{\overline{Z}}^{RS} \, \overline{\chi}_{CDR} \kappa_S
    + \mbox{Hermitian conjugate\ ,}
\end{eqnarray}
where special attention must be paid to some details.
In the first line of (\ref{puffo}) the
  term $\overline{\chi}_{[CDR} \,\kappa_{S]}$ is completely
  antisymmetrized in the indices $CDRS$, while the same term in
  the second line of equation~(\ref{puffo}) is not
  antisymmetrized. Moreover, some of the central charge tensors appearing in
  equation~(\ref{puffo}) wear a hat, namely are projected, and some
  do not.
\par
We can now discuss the cancellation of the dilatino terms for the
three values of $\nu$.
\begin{description}
  \item[{\underline{Case $2\nu =8$}}. ]{ Here the projection operator
  ${\cal P}^A_B$ is the identity operator and we have
  $Z_{AB}=\widehat{Z}_{AB}$. Furthermore, in view of the identity (\ref{idezz}) we
  see that the  two terms on the first line of (\ref{puffo}) are identical and sum together.
  On the other hand, we also have
  $\overline{Z}^{[CD}\,\overline{Z}^{R]S} =\overline{Z}^{[CD}\,
  \overline{Z}^{RS]}$  so that the cancellation of the dilatino terms occurs if:
\begin{equation}\label{dilcan8}
  \alpha_{8}=\ft{1}{2}\sqrt{2}\ ,
\end{equation}
which for $\nu =4$ coincides with (\ref{alpcof}).}
  \item[{\underline{Case $2\nu =4$}}. ] Here there is a difference
  between $\widehat{Z}_{AB}$ and $Z_{AB}$.  A little analysis
 is needed to show that the first and second term in the first line
  together reconstruct indeed all the possible cases
  present in the second line of (\ref{puffo}), upon use of the
  extension of the identity (\ref{idezz}) to the case that
  one of the central charges is projected.
One eventually finds that, in agreement with (\ref{alpcof}),
 the dilatino terms cancel if
\begin{equation}\label{dilcan4}
  \alpha_{4}=1\ .
\end{equation}
  \item[{\underline{Case $2\nu =2$}}. ] {In this last case the
  second term in the first line of equation~(\ref{puffo}) vanishes
  identically since there is antisymmetrization on the indices
  $IJ$ and $S$ that all lie in a two-dimensional subspace. As for
  the first term on the same line, in order to single out the
  non-vanishing contribution that can cancel with the similar term
  in the second line one has to develop the antisymmetrization.
  We have
\begin{equation}\label{undosym}
  \widehat{\overline{{Z}}}^{CD} \,\overline{Z}^{RS}\,
  \overline{\chi}_{[CDR} \,\kappa_{S]}=\ft{1}{2}
  \widehat{\overline{{Z}}}^{CD} \,\overline{Z}^{RS}\,
  \overline{\chi}_{ CDR} \,\kappa_{S}+\ft{1}{2}
  \overline{Z}^{CD}\,\widehat{\overline{{Z}}}^{RS} \,
  \overline{\chi}_{ CDR} \,\kappa_{S}\ ,
\end{equation}
and the first of the two terms appearing on the right hand side of
(\ref{undosym}), vanishes for the same reason as before. We cannot
antisymmetrize in $CDR$ when all of the three indices lie in the
two-dimensional subspace singled out by the projection. On the
other hand, the second term on the right-hand side of
(\ref{undosym}) can cancel with the second line of equation~(\ref{puffo}).
For this case, therefore, the dilatino terms
cancels if
\begin{equation}\label{dilcan2}
  \alpha_2  = \sqrt{2}\ ,
\end{equation}
which is once again consistent with the $\nu=2$ case of
(\ref{alpcof}).}
\end{description}
In this way we have completed the proof of $\kappa$-supersymmetry
and shown that   (\ref{0actn8}) is the correct $0$-brane action
for all ${\cal N}=8$ supergravity backgrounds.
\section{Outlook}
In this paper we have constructed $0$-brane actions for
superparticles moving in generic $D=4$ supergravity backgrounds
that preserve a residual fraction of supersymmetry. Typically such
backgrounds are BPS black holes. The main issue in writing these
actions is the coupling of the $0$-brane, not only to the metric and
the gauge fields of the background, but also to its scalar fields.
\par
The main application of our result appears to be  its possible use
as an instrument to investigate the structure and the properties
of $(1+0)$-dimensional conformal field theories (superconformal
quantum mechanics) living on the boundary of two-dimensional
anti-de~Sitter space. Indeed, in those cases where the black hole
entropy is finite,  the near-horizon geometry of the black hole
is $adS_2 \times S^2$ and we can formulate the Kaluza--Klein
programme for $D=4$ supergravity as its compactification on $S^2$.
In complete analogy with \cite{termoniaetal} we can then study the
conformal field theory on the $adS_2$ boundary starting from our
superparticle action. Other applications are possible but this is
the main one that motivated us to undertake the present study.
\par
In deriving our result, one main point that had to be cleared, was
the correct formulation of the projection equation satisfied by
the $\kappa$-supersymmetry parameter. This projection is
identical to the equation satisfied by the BPS Killing spinor
admitted by the  corresponding supergravity background. In
previous literature this Killing spinor equation was written in
fixed $U({\cal N})$ gauges. In the present paper we have restored
its complete $U({\cal N})$ covariance, $SU(8)$ covariance for
${\cal N}=8$.
\medskip
\section*{Acknowledgments.}
\noindent This work was supported by the European Commission TMR
programme ERBFMRX-CT96-0045, in which S.C. and D.Z. are associated
with the University of Torino.
\newpage
\appendix
\section{Notations and self-duality of antisymmetric tensors}  \label{app:signsforms}
In table~\ref{tbl:compind} we give a list of indices, listing its range, and
comparing with notations in some of the references.
\begin{table}[ht]\caption{Comparison of indices and normalizations}
\label{tbl:compind}\begin{center}
\begin{tabular}{|lllll|}\hline
Here &
\cite{DWVP,dWLVP,trsummer}&\cite{n2stand,RiccardoPietroReview,bholIt1,bholIt2,bholIt3}
& Meaning & Range \\ \hline
$i,\ib$  & $\alpha,\bar\alpha$ & $i,i^*$ & Moduli & $1,\ldots, n$ \\
$\ILambda$  & $I$ & $\Lambda$ & Symplectic & $0,\ldots, n$ \\
$A$  & $i$ & $A$ & Extended susy & $1,2$ \\
$\alpha,\bar\alpha$  &  & $\alpha,\bar\alpha$ & CY coordinates & $1,2,3$ \\
$\mu$  & $\mu$ &$\mu$  & Local $D=4$  & $0,\ldots,3$ \\
$a$  & $a$ & $a$ & flat $D=4$ & $0,\ldots,3$ \\
$i$  &  &  & Space  & $1,2,3$ \\
$M$  &  &  & $D=10$ local & $0,\ldots,9$ \\
$m$  &  &  & $D=4$ world-volume & $0,\ldots,3$ \\ \hline
$g_{\mu\nu}$ & $g_{\mu\nu}$ & $-g_{\mu\nu}$ & Metric & \\
$\mathcal{F}_{\mu\nu}$ & $-\mathcal{F}_{\mu\nu}$
&2$\mathcal{F}_{\mu\nu}$ & Field strengths & \\
$A_\mu$ &$-W_\mu$ &$A_\mu$ & Vectors & \\
$\epsilon_{\mu \nu\rho\sigma}$ & $-\ii\epsilon_{\mu \nu\rho\sigma}$ &
 $\epsilon_{\mu \nu\rho\sigma}$ &Levi-Civita tensor & \\
$\gamma^a$ & $\gamma^a$ & $\ii\gamma^a$ & & \\
$\XL^\ILambda$ & $X^I$ & $L^\Lambda$ & Upper cov. hol. section & \\
$\FM_\ILambda$ & $F_I$ & $M_\Lambda$ & Lower cov. hol. section & \\
& $Z^I$ & $X^\Lambda$ & Upper   hol. section & \\
&  & $F_\Lambda$ & Lower  hol. section & \\
$T_{\mu \nu }$ &$ \ft i4 T^{-ij}_{\mu \nu }\epsilon_{ij}$ & $2T_{\mu \nu
}$& Gravitino field strength & \\
$\epsilon_A$ & $\sqrt{2}\epsilon^i$ &$ \epsilon_A$ & susy parameter & \\
$\psi_\mu^A$&$ \frac{1}{\sqrt{2}}\psi_{\mu \,i}$ &$\psi_\mu^A$ &
gravitino & \\
\hline
\end{tabular}\end{center}\end{table}
Our spacetime metric is thus mostly $+$, which is opposite to
\cite{n2stand,RiccardoPietroReview,bholIt1,bholIt2,bholIt3}. There are more
small differences in normalization between some of the articles in
the same group. In the table we use the most recent ones. A more
detailed comparison can be found in the appendix~B of
\cite{trsummer}. For $\mathcal{N}=8$, of course, $A=1,\ldots, 8$, and we use also $\Lambda
,\Sigma =1,\ldots, 8$, related to $E_7$.
\par
We adopt the notation where in any dimension $\epsilon_{01\ldots
(d-1)}=1$, and the dual $\tilde F$ of a tensor $F$ is defined such that it squares to
the identity: $\tilde{\!{\tilde F}}=F$. With complex variables in the CY
(say $y^\alpha$ and $\bar y^{\bar\alpha}$),
we have
\begin{equation}
\epsilon_{\mu \nu\rho\sigma \alpha\beta\gamma\bar\alpha\bar\beta\bar \gamma}=
\ii\epsilon_{\mu \nu\rho\sigma}
\epsilon_{\alpha\beta\gamma} \epsilon_{\bar\alpha\bar\beta\bar \gamma}\ ,
   \label{eps10in433}
\end{equation}
where the $\ii$ now appears by going from real indices $5,\ldots,9$
to holomorphic and anti-holomorphic indices $\alpha,\bar\alpha$.
Duality in 10-dimensional Minkowski space should be defined by
\begin{equation}
\tilde F_{MNPQR}=\frac{1}{5!}\epsilon_{MNPQRSTUVW}F^{STUVW} \ ,
\end{equation}
in order that $\tilde{\!{\tilde F}}=F$.
Taking a 5-form with only non-zero components
$F_{\mu \nu \alpha\beta\gamma}=F_{\mu \nu}\Omega_{\alpha\beta\gamma}$, and taking the
metric with $g^{\alpha\bar\alpha}=1$, the tensor with upper indices has non-zero
components $F^{\mu \nu \bar\alpha\bar\beta\bar \gamma}=F^{\mu \nu}\Omega_{\alpha\beta\gamma}$.
The factor $\ii$ in (\ref{eps10in433})  is absorbed in taking the dual in
4 dimensions, such that we have
\begin{equation}
F_{\mu \nu \alpha\beta\gamma}=F_{\mu \nu}\Omega^{(3,0)}_{\alpha\beta\gamma}
\quad \rightarrow\quad
\tilde F_{\mu \nu \alpha\beta\gamma}=\tilde F_{\mu \nu} \Omega^{(3,0)}_{\alpha\beta\gamma}  \ .
\end{equation}
For the other 3-forms in the CY some re-ordering of indices has to
be made in the Levi-Civita tensor, such that
\begin{eqnarray}
F_{\mu \nu \alpha\beta\bar \gamma}&=&F_{\mu \nu}\Omega^{(2,1)}_{\alpha\beta\bar\gamma}
\quad \rightarrow\quad
\tilde F_{\mu \nu \alpha\beta\bar \gamma}=-\tilde F_{\mu \nu}
 \Omega^{(2,1)}_{\alpha\beta\bar\gamma}
\nonumber\\
F_{\mu \nu \alpha\bar\beta\bar \gamma}&=&
F_{\mu \nu}\Omega^{(1,2)}_{i\bar\beta\bar\gamma}\quad \rightarrow\quad
\tilde F_{\mu \nu \alpha\bar\beta\bar \gamma}=
\tilde F_{\mu \nu} \Omega^{(1,2)}_{\alpha\bar\beta\bar \gamma}
\nonumber\\
F_{\mu \nu \bar\alpha\bar\beta\bar\gamma}&=&
F_{\mu \nu}\Omega^{(0,3)}_{\bar\alpha\bar\beta\bar\gamma}\quad \rightarrow\quad
\tilde F_{\mu \nu \bar\alpha\bar\beta\bar\gamma}=-\tilde F_{\mu \nu}
\Omega^{(0,3)}_{\bar\alpha\bar\beta\bar\gamma}  \ .
\label{dualin10}
\end{eqnarray}



\begin{thebibliography}{99}
\bibitem{Maldacena} J. Maldacena, {\it The Large N limit of superconformal
field theories and supergravity},
Adv. Theor. Math. Phys. {\bf 2} (1997)  231;
hep-th/9711200.
\bibitem{GT} G. W. Gibbons and P. K. Townsend,
{\it Vacuum interpolation in supergravity via super p-branes},
Phys. Rev. Letters {\bf 71} (1993) 3754; hep-th/9307049.
\bibitem{stellelectur} K.S. Stelle, {\it Lectures on supergravity
$p$-branes}, 1996 ICTP Summer School in High Energy Physics and Cosmology, Trieste,
eds. E. Gava, A. Masiero, K. Narain, S. Randjbar-Daemi and Q. Shafi,
World Scientific, p. 287; hep-th/9701088.
\bibitem{kappa} M.T. Grisaru, P. Howe, L. Mezincescu, B. Nilsson and
P.K. Townsend, {\it $N=2$ superstrings in a supergravity background},
Phys. Lett. {\bf 162B} (1985) 116;\\
P.K. Townsend, {\it Spacetime supersymmetric particles and strings in
background fields}, in `Superunification and extra dimensions', eds.
R. D'Auria and P. Fr\`e, (World Scientific, 1986), p. 376;\\
M. Tonin, {\it Consistency condition for kappa anomalies and superspace constraints in quantum heterotic
superstrings}, Int.J.Mod.Phys. {\bf A4} (1989) 1983.
\bibitem{castdauriafre} L. Castellani, R. D'Auria and P. Fr\`e, {\em
Supergravity and Superstring theory: a geometric perspective}, World
Scientific, 1990.
\bibitem{termoniaetal} G. Dall'Agata, D. Fabbri, C. Fraser, P. Fr\`e, P. Termonia
and M. Trigiante, {\it The $OSp(8|4)$ singleton action from the
supermembrane}, Nucl. Phys. {\bf B542} (1999) 157; hep-th/9807115.
\bibitem{conffadS} P. Claus, R. Kallosh and A. Van Proeyen,
{\it M 5-brane and superconformal (0,2) tensor multiplet in 6
dimensions}, Nucl. Phys. {\bf B518} (1998) 117, hep-th/9711161\\
 P. Claus, R. Kallosh, J. Kumar, P.K. Townsend and
A. Van Proeyen, {\it  Conformal theory of M2, D3, M5 and `D1+D5' branes},
JHEP {\bf 06} (1998) 004, hep-th/9801206.
\bibitem{adS5S5} R.R. Metsaev and A.A. Tseytlin, {\it Type IIB superstring action in $AdS_5 \times
S^5$ background}, Nucl. Phys. {\bf B533} (1998) 109; hep-th/9805028;\\
 I. Pesando,
{\it A kappa fixed type~IIB superstring action on $AdS_5 \times
S^5$}, JHEP {\bf 11} (1998) 002; hep-th/9808020;\\
R. Kallosh, J. Rahmfeld, {\it The GS string action on $AdS_5 \times
S^5$}, Phys. Lett. {\bf B443} (1998) 143; hep-th/9808038.
\bibitem{GH} G.W. Gibbons and C.M. Hull, {\em A Bogomolny bound for general
relativity and solitons in N=2 Supergravity}, Phys. Lett.  {\bf
109B},  190  (1982);\\
 G.W. Gibbons, in: {\it Supersymmetry, Supergravity and
Related Topics},  eds. F. del Aguila, J. de Azc\'arraga and L.
Ib\'a\~nez
(World Scientific,  Singapore 1985), p. 147;\\
 K.P. Tod, {\em All metrics admitting supercovariantly constant spinors},
 Phys. Lett. {\bf 121B}, 241 (1983).
\bibitem{FKS} S. Ferrara, R. Kallosh and A. Strominger, {\em N=2
extremal black holes}, Phys. Rev.
{\bf D52} (1995) 5412.
\bibitem{RiccardoPietroReview} R. D'Auria and P.Fr\`e, {\em BPS Black Holes in
Supergravity}, Lecture Notes for the SIGRAV Graduate School in Contemporary
Relativity, Villa Olmo, Como First Course, April 1998;
hep-th/9812160.
\bibitem{bhscmech} P. Claus, M. Derix, R. Kallosh, J. Kumar, P. K.
Townsend and A. Van Proeyen, {\it Black Holes and Superconformal
Mechanics}, Phys. Rev. Lett. {\bf 81} (1998) 4553; hep-th/9808038.
\bibitem{SergioRenataGary} S. Ferrara and R. Kallosh, {\it
Supersymmetry and Attractors}, Phys. Rev.\textbf{D54} (1996) 1514,
hep-th/9602136;\\
S. Ferrara, G.W. Gibbons and R. Kallosh, {\it
 Black Holes and critical points in moduli space}, Nucl. Phys. \textbf{B500}
 (1997) 75, hep-th/9702103.
\bibitem{berrobin} B. Bertotti, Phys. Rev. {\bf 116} (1959) 1331;\\
I. Robinson, Bull. Acad. Pol. {\bf 7} (1959) 351
\bibitem{DWVP}
B. de Wit and A. Van Proeyen, {\it Potential and symmetries of general gauged
$N=2$ supergravity-Yang-Mills models}, Nucl. Phys. {\bf B245} (1984) 89.
\bibitem{n2stand} L. Andrianopoli, M. Bertolini, A. Ceresole, R. D'Auria,
S. Ferrara, P. Fr\'e and T. Magri, {\em N=2 supergravity and N=2 super Yang-Mills
theory on general scalar manifolds:
symplectic covariance, gaugings and the momentum map}, Jour. Geom and Phys.
{\bf 23} (1997) 111; hep-th/9605032.
\bibitem{Moore} G. Moore, {\em Arithmetic and attractors},
{\tt hep-th/9807087}.
\bibitem{FredAttractor} F. Denef, {\it Attractors at weak gravity},
 Nucl. Phys. {\bf B547} (1999) 201; hep-th/9812049.
\bibitem{dWLVP} B. de Wit, P. Lauwers and A. Van Proeyen, {\it Lagrangians
of $N=2$ supergravity-matter systems}, Nucl. Phys. {\bf B255} (1985) 569.
\bibitem{BBS}K.
Becker, M. Becker and A. Strominger, {\em Fivebranes, Membranes and
Non-Perturbative String Theory}, Nucl.Phys. B456 (1995) 130,
hep-th/9507158
\bibitem{Polchinski2} J. Polchinski, {\em String theory}, Vol.II,
Cambridge Univ. Press, 1998.
\bibitem{Iengo} M. Bertolini, P. Fr\`e, R. Iengo, C.A. Scrucca,
 \textit{Black holes as D3-branes on Calabi--Yau threefolds}, Phys. Lett. {\bf B431} (1998)
 22.
\bibitem{bholIt1} L. Andrianopoli, R. D'Auria, S. Ferrara, P. Fr\'e,
and M. Trigiante, {\it $E_{7(7)}$ Duality, BPS Black-Hole Evolution and
 Fixed Scalars}, Nucl. Phys. {\bf B509} (1998) 463, hep-th/9707087.
\bibitem{bholIt2} G. Arcioni, A. Ceresole, F. Cordaro, R. D'Auria, P.
Fr\'e, L. Gualtieri and M. Trigiante, {\it $N=8$ BPS black holes with 1/2 or
1/4 supersymmetry and solvable Lie algebra decompositions}, Nucl. Phys. {\bf B542} (1999)
273; hep-th/9807136.
\bibitem{bholIt3} M. Bertolini, P. Fr\'e, M. Trigiante, {\it $N=8$ BPS black holes preserving
1/8 supersymmetry}, Class. Quantum Grav. {\bf 16} (1999) 1519,  hep-th/9811251.
\bibitem{myneucha} P. Fr\'e, {\em Solvable Lie algebras, BPS black holes
 and supergravity gaugings}, Fortsch. Phys. {\bf 47} (1999) 173;
hep-th/9802045
\bibitem{cre}E. Cremmer, B. Julia, {The $N=8$ supergravity theory.
 1. the Lagrangian}, Phys.Lett. {\bf 80B} (1978) 48; {\it The $SO(8)$
 supergravity}, Nucl.Phys. {\bf B159} (1979) 141.
\bibitem{LauSerRiczent} L. Andrianopoli, R. D'Auria and  S. Ferrara,
{\em $U$ duality and central charges in various dimensions
revisited}, Int. J. Mod. Phys. {\bf A13} (1998) 431; hep-th/9612105.
\bibitem{trsummer} A. Van Proeyen, {\it Vector multiplets in $N=2$ supersymmetry
 and its associated moduli spaces},
in  {\it
1995 Summer school in High Energy Physics and Cosmology}, The ICTP
series in theoretical physics - vol.12 (World Scientific, 1997), eds.
E. Gava et al., p.256;
hep-th/9512139.
\end{thebibliography}
\end{document}